\documentstyle[preprint,eqsecnum,aps]{revtex}

\newcommand{\be}{\begin{equation}}
\newcommand{\ee}{\end{equation}}
\newcommand{\bea}{\begin{eqnarray}}
\newcommand{\eea}{\end{eqnarray}}

\newcommand{\fsl}{\not\!}

\renewcommand{\baselinestretch}{1.0}
\begin{document}

\preprint{HD-TVP-98/02}
\title{\vspace{1.5cm} Transition rates for $
q\bar q\rightarrow \pi\pi\pi$ in a
chiral model}
\author{D.~S. Isert, S.~P.~Klevansky and P. Rehberg\footnote{Current address:
SUBATECH, Laboratoire de Physique Subatomique et des
         Technologies Associ\'ees
         UMR, Universit\'e de Nantes, IN2P3/CNRS, Ecole des Mines de Nantes
         4 Rue Alfred Kastler, F-44070 Nantes Cedex 03, France.
} }
\address{ Institut f\"ur Theoretische Physik, \\
Philosophenweg 19, D-69120 Heidelberg, Germany}

\maketitle
\vspace{1cm}

\begin{abstract}
We investigate the nature of
transition rates for the hadronization process of $q\bar q\rightarrow\pi\pi\pi$
as opposed to the transition rates for $q\bar q\rightarrow\pi\pi$,
within the Nambu--Jona-Lasinio model that has manifest chiral symmetry.
Feynman diagrams appropriate to this process can be classified according
to the expansion in the inverse number of colors $1/N_c$.    Two of these
types of graphs are seen to be either ``$s$-like'' or ``$t$-like'' in nature.
A further graph that contains both $s$-like and $t$-like elements, and which
is denoted as $st$-like, is also present.   To
describe such a process with two incoming and three outgoing particles,
it is necessary to extend the number of Mandelstam invariants.   It is
convenient to introduce
 seven such variables, of
which only five are independent.
  The cross
section for two incoming particles to three outgoing particles is
then
reexpressed
in integral form in terms of these invariants.   As a function of $\sqrt s$,
the final expression is reduced to
 an integral over the four remaining invariants.
The limits of integration,
which are now non-trivial, are also discussed.   The transition rate for
 the explicit case of $u\bar u\rightarrow \pi^+\pi^-\pi^0$, is evaluated
numerically, assuming non-chiral pions, $m_\pi=135$MeV.   The rate for three
pion production is found to be of the same order
of magnitude as for two pion production,
making this a non-negligible contribution to hadronization.
\end{abstract}
\clearpage

\renewcommand{\baselinestretch}{1.3}
\footnotesize\normalsize

\section{Introduction}
Nonequilibrium formulations of field theories are in a stage of development
at present.   Their study is essential for ultimately constructing a solid
theoretical basis for the transport codes that are used according to various
models for the simulation of heavy-ion collisions.   One such attempt
that is still in its early stages investigates the nonequilibrium
formulation of a {\it chiral} theory, the Nambu--Jona-Lasinio (NJL) model
\cite{nambu,reviews}.
{}From practical experience, one assumes that a collision term of the
associated semi-classical limit will be Boltzmann-like in form, and this
has been demonstrated explicitly \cite{ogu}.   In particular, a full
calculation
that includes mesonic degrees of freedom must also contain not only
quark-quark and quark-antiquark scattering cross-sections on the partonic
level, but also the hadronization processes $q\bar q\rightarrow MM'$
and their dissolution, where
$M$ represents a mesonic state.    Cross-sections for these processes in
SU(3) have already been calculated and are given elsewhere \cite{pr}.

Processes such as those described in the last paragraph are binary in
nature.   On the other hand, there are numerous other processes that may
occur and 
which  increase the total number of particles. While binary collisions lead
to thermal equilibrium, other multiparticle
collisions may also be
 required to attain chemical equilibrium.   In particular,
the simplest possible further kind of collision would be one that increases
the number of particles, 
 such as that of $q\bar q\rightarrow MM'M''$.   Such a process should 
also rightly  enter into the collision dynamics. 
If one had some knowledge of the magnitude of such processes, one would
at least know whether multiparticle production in transport models could be
expected to play an important role or not. 
   In particular, one imagines that multiparticle hadronization processes
such as the type just mentioned 
could be particularly 
important for the Goldstone bosons, the pions, since they  are
 massless in the chiral limit, and are light in any event.
Since pions are the dominant degree of freedom in the late phase of 
heavy-ion collisions, being produced copiously, it is essential to study
their role.
  The purpose of this paper is thus to
examine the cross-section for processes of the kind
$q\bar q\rightarrow \pi\pi\pi$, where two particles
enter the interaction region, while three are emitted.   A direct numerical
calculation is performed specifically for the case $u\bar u\rightarrow
\pi^+\pi^-\pi^0$.   This specific example has been chosen because the
hadronization channel into two mesons $u\bar u\rightarrow \pi^+\pi^-$
has already been calculated and has been shown to
dominate over the other possible existing two pseudoscalar emission
 channels that include
 $u\bar u\rightarrow \pi^0\pi^0$, $K^+K^-$ and $\pi^0\eta$ \cite{pr}.

The chiral model underpinning our calculation is the NJL
model, in which      the SU(2) version is used.
While this model has several shortcomings, specifically that it does not
confine, it gives an excellent description of the mesonic and baryonic
sectors \cite{reviews}.   In using this model, one is able to obtain a 
self-consistent description within a chiral theory,
 which ultimately may be used in developing a
self-consistent transport theory that in turn is 
based purely on this chiral Lagrangian 
\cite{ogu,aich,ph}.

  As a guide to determining the most relevant diagrams,
the expansion in the inverse number of colors $1/N_c$ is used
\cite{quack,dmitra}.  Within the same order of $1/N_c$, there are three
types of diagrams, an $s$-like, a $t$-like, and a mixed so-called
$st$-like graph.   We find expressions for the invariant amplitudes in all
three cases, and apply this to a calculation of the transition rates
specifically for the process $u\bar u\rightarrow \pi^+\pi^-\pi^0$.
Since there are many terms that contribute to each channel, we investigate
all channels first separately, and find that the $st$-like channel is at
least an order of magnitude larger than the $s$-like and $t$-like graphs.
That is to say, within the $1/N_c$ ordering, there appears to be a 
subordering that dictates that one (or more) graphs are in fact dominant.
In retrospect, one can give {\it heuristic} arguments why the
particular $st$-like graph should in fact
dominate.  This particular graph contains the most divergent functions,
since it contains a fermion loop with three propagators.
The $s$-like graph by contrast contains a fermion loop with four
propagators. In fact, such an
ordering of terms within one specific class of diagrams that superficially
are all of the same order in $1/N_c$ {\it rather} 
according to the degree of divergence of each graph was proposed long
ago in Ref.\cite{eguchi}. The calculation in this paper thus serves
also in part to
validate {\it a posteriori}
that this ordering correctly assesses the importance of each term.   In
Ref.\cite{eguchi}, convergent terms were regarded as being secondary, while
the
leading behavior was determined solely by the divergent terms.
In a further calculation of the pion radius \cite{hippe}, it has also
been seen that the leading logarithmic contribution that is in accord
with the well-known results of sigma models and 
chiral perturbation theory, is also
due to the most divergent graph of the many available to the same
order in the $1/N_c$ expansion  when calculated in the NJL model. 
This is so because this graph naturally corresponds to the pion loop diagram
that has been known to give rise to the chiral logarithm in so many other
models or theories, see for example Refs.~\cite{tarrach,gasser,koffer}.
Note that in Ref.~\cite{hippe},
 however, no explicit calculations of the remaining terms
was undertaken.

   In our numerical
calculation of the transition rate that was carried out using a Monte
Carlo procedure, we find that the $st$-like channel has the same order of
magnitude for the production rate of $u\bar u\rightarrow\pi^+\pi^-\pi^0$
as is found for the dominant two particle production rate $u\bar u
\rightarrow \pi^+\pi^-$ \cite{pr}.   As such, it appears to give a
nonnegligible contribution to the production cross-section for three pions
and
cannot be ignored in hadronization programs at low energies.

An essential feature of the calculation is the construction of the
cross-section for the process of two incoming to three outgoing states,
$a + b \rightarrow c + d + e$.     To this end, we introduce extended
Mandelstam-like variables that will be denoted as $s,t,u,v,w,x$ and $y$,
and we reexpress the cross-section in terms of these variables.    The
invariant amplitude, which is given by the relevant Feynman graphs is
most easily expressible in terms of these extended variables
\footnote{Other choices are also possible \cite{buck}.  The variables
chosen here however, arise naturally when one examines the possible
$t$-channel processes.}.
Although this section of our work is performed for the case
of incoming quarks and outgoing pions, and has thus the restriction
$p^2_1=p^2_2=m^2$ and $p^2_3=p^2_4=p_5^2=m_\pi^2$, the kinematics are
otherwise unrestricted so that the generalization to other arbitrary
three body processes is obvious.

This paper is organised as follows.
  In Section II, we
evaluate the invariant amplitude for two arbitrary incoming quark flavors and
outgoing mesonic states.   In Section III, we discuss the
kinematics
of the process $a+b\rightarrow c+d+e$ and write the cross-section for this
process in terms of extended Mandelstam-like variables.  In Section IV, we
calculate the cross-section
for $u\bar u\rightarrow \pi^+\pi^-\pi^0$, and present our numerical results.
We summarize and conclude in Section V.

\section{Hadronization into three pions.}

In this section, we discuss the processes that lead to three pion
production, governed by the SU(2) Nambu--Jona-Lasinio Lagrangian density
\begin{equation}
{\cal L} = \bar \psi(i\!\not\!\partial - m_0)\psi + G[(\bar\psi\psi)^2 + (\bar
\psi i\gamma_5\vec\tau\psi)^2] \ .
\label{e:lag}
\end{equation}
Here $m_0^u = m_0^d = m_0$ is the current quark mass for $u$ and $d$ quarks
and
$G$ is a dimensionful coupling strength, $[G]$ = [MeV]$^{-2}$.   There are
several detailed reviews of this model in the literature, see for example
Refs.\cite{reviews}.    The reader is referred to these references for a
derivation of
the  gap equation that determines the constituent quark mass
\begin{equation}
m=m_0 + 2iGN_cN_f\int\frac{d^4p}{(2\pi)^4} {\rm tr} S(p)
\label{e:gap}
\end{equation}
with $S(p)^{-1} = \!\not\!\!p-m$ as well as for obtaining
 the pseudoscalar and scalar meson masses via the condition
\begin{equation}
1-2G\Pi_{S,PS}(k^2)|_{m_\sigma^2,m_\pi^2} = 0,
\label{e:det}
\end{equation}
where the scalar and pseudoscalar polarization functions are defined in the
usual way as
\begin{eqnarray}
\Pi_{PS}(k^2) &=& -iN_c {\rm tr}_f(T_iT_f) \int\frac{d^4p}{(2\pi)^4}\frac
{(-\!\not\! p + m)(\!\not\! p+\!\not\! k + m)}{[p^2-m^2][(p+k)^2-m^2]} \\
\Pi_{S}(k^2) &=& iN_c  \int\frac{d^4p}{(2\pi)^4}\frac
{(\!\not\! p + m)(\!\not\! p+\!\not\! k + m)}{[p^2-m^2][(p+k)^2-m^2]}
\label{e:pols}
\end{eqnarray}
with $T_i$ an isospin operator $T_3=\tau_3$ and $T_{\pm} = (
\tau_1\pm i\tau_2)/\sqrt2$.

Using the concept now that one can construct the Feynman diagrams for
any process involving quarks, antiquarks and mesons, one can classify the
graphs that contribute to hadronization.   In the case that one has
hadronization into {\it two} particles, this is given in the standard
fashion according to $s$ channel, $t$ channel and $u$ channel graphs, as
is shown in Fig. 1.   Here $s$, $t$ and $u$ are the usual Mandelstam
variables, $s=(p_1+p_2)^2$, $t=(p_1-p_3)^2$ and $u=(p_1-p_4)^2$.    From
energy-momentum conservation, $p_1+p_2=p_3+p_4$, one has the additional
constraint
\begin{equation}
s + t + u = 2m^2 + 2m_\pi^2.
\label{e:sum}
\end{equation}
These amplitudes and their resulting cross-sections have been discussed
in detail in Ref.\cite{pr}.   Note that the set of graphs is chosen to be
leading
in the $1/N_c$ expansion \cite{quack,dmitra}.
    That the $s$, $t$ and $u$ channels are of
the same order lies in the fact that  meson propagators (in the
pionic channel) are constructed as intermediate states into 
quark-antiquark scattering amplitudes.    These quark-antiquark scattering
amplitudes contribute as
$g^2_{\pi q q}/m_\pi^2\sim 1/N_c$, 
as has been explicitly shown in Ref.~\cite{dmitra}, for example.
A fermionic loop on the other hand brings a contribution of order $\sim N_c$.

   Generic diagrams that we  can construct  for two incoming quarks
and three outgoing mesons are shown
in Fig.~2(a)-(c).   We note that in direct analogy to the graphs of Fig.~1,
we can order the hadronization graphs of Figs.~2(a) and 2(b) as
being $s$-{\it like} or
$t$-{\it like}.  Figure 2(c) fits into neither category and we thus denote
it as $st$-{\it like}.    As has already been stated, the graphs of Fig.~2
are
generic:
   the full set of diagrams is obtained for Figs.~2(a) and 2(b)
by permuting the outgoing
momenta.
There are  $3! =6$ ways of doing this.
Whether all 6 diagrams in each case
contribute to a process or not is then  determined by the
flavor factors of the individual incoming quarks in question.
In Fig.~2(c), four cases arise on noting that the vertex for two particle
production can be attached either to the incoming (top or bottom) quark or
antiquark line, and the sense of the line in the quark loop can be clockwise
or anticlockwise.  The single emanating meson line can also take on any of
the final state momenta, leading to a further permutation of these three
possibilities.
    Exact combinations again depend on the flavors of the
vertices in question.

For two incoming states to three outgoing states, it is useful to define the
extended Mandelstam-like variables,
\begin{eqnarray}
s &=& (p_1+p_2)^2 \nonumber\\
t &=& (p_1-p_3)^2 \nonumber \\
u &=& (p_1-p_4)^2 \nonumber \\
v &=& (p_1-p_5)^2 \nonumber \\
w &=& (p_1-p_3-p_4)^2 = (p_2-p_5)^2 \label{e:exmand} \\
x &=& (p_1-p_4-p_5)^2 = (p_2-p_3)^2 \nonumber \\
y &=& (p_1-p_3-p_5)^2 = (p_2 -p_4)^2, \nonumber
\end{eqnarray}
where the last relations follow from
energy-momentum conservation, $p_1+p_2=p_3+p_4+p_5$.
Using, in addition, the restriction to quarks for the incoming states,
\begin{equation}
p_1^2=p_2^2=m^2
\end{equation}
and pions for the outgoing states,
\begin{equation}
p_3^2=p_4^2=p_5^2=m_\pi^2,
\label{e:outs}
\end{equation}
we find
\begin{equation}
s+t+u +v = 3m^2 + 3m_\pi^2
\label{e:newrel1}
\end{equation}
and
\begin{equation}
s+w+x+y = 3m^2+3m_\pi^2,
\label{e:newrel2}
\end{equation}
in analogy to Eq.(\ref{e:sum}).

In what follows, in our analysis of the  $s$-like and  $t$-like graphs,
 the products $p_1p_2$, $p_1p_3$ and
so forth, occur.   These are simply related to the extended Mandelstam
variables via, e.\,g.,
\begin{eqnarray}
p_1\cdot p_2&=& \frac 12 (s-2m^2) \nonumber \\
p_1\cdot p_3&=& -\frac12(t-m^2-m_\pi^2).
\label{e:invert}
\end{eqnarray}
We do not explicitly rewrite our expressions in terms of these variables:
the numerical program makes a direct translation.  In contrast to this, the
$t$-like channel amplitudes are most simply written directly in terms of
these variables and we thus use them  explicitly from the start.

\subsection{$s$-like channel}

In this section,  we discuss the $s$-like channel diagram involving an
intermediate scalar or pseudoscalar meson.   One of the six possible
ways of distributing the momenta is shown in Fig.~3.
We shall assume that the final state always contains three pseudoscalar
particles.    In this case, if the intermediate meson is a scalar,
the scattering amplitude, vanishes,
\begin{equation}
-i{\cal M}_{\sigma}(p_3,p_4,p_5)=0.
\label{e:msig}
\end{equation}
This must be so on general grounds, since the parity of the final state is
negative, in contrast to the positive parity of the intermediate state.
This can also be validated by direct calculation.

We thus require Fig.~3 calculated with  the pion in the intermediate state.
We examine this single diagram first with the momenta as indicated in the
figure.    Using an obvious notation, this is
\begin{eqnarray}
-i{\cal M}_{\pi
}(p_3,p_4,p_5) &=& \bar v_2 i\gamma_5T_ju_1\delta_{c_1c_2}
\frac{2iG}{1-2G\Pi_{PS}(k^2)} (ig_{\pi q q})^3
A_{\pi\pi\pi\pi}(p_3,p_4,p_5), \nonumber \\
\label{e:mpi}
\end{eqnarray}
where the four pion vertex is given as
\begin{eqnarray}
&&A_{\pi\pi\pi\pi}(p_3,p_4,p_5) = \nonumber \\
&& -{\rm Tr}\int\frac{d^4q}{(2\pi)^4} i\gamma_5
T_kiS(q-k)i\gamma_5 T_5 iS(q-p_3-p_4) i\gamma_5 T_4 iS(q-p_3) i \gamma_5 T_3
iS(q).\nonumber \\
\label{e:a}
\end{eqnarray}
In this expression, ${\rm Tr} = {\rm tr}_\gamma {\rm tr}_f {\rm
tr}_c$ sums over all degrees of
freedom, and $T_i$ is an appropriate flavor operator.   Color appears
trivially as shown by the $\delta$-function in Eq.(\ref{e:mpi}).
Inserting
the single particle propagator leads to the form
\begin{eqnarray}
&&A_{\pi\pi\pi\pi}(p_3,p_4,p_5) =-N_cN_f^{(345k)}
\times \nonumber \\
&&\int\frac{d^4q}{(2\pi)^4}
\frac{{\rm tr}_\gamma[(-\!\not\! q + \!\not\! k + m)(\!\not\!
 q - \!\not\! p_3-\!\not\! p_4+m)(-\!\not\! q + \!\not\!
p_3 + m) (\!\not\! q + m)]}{(q^2-m^2)[(q-p_3)^2
-m^2][(q-p_3-p_4)^2-m^2][(q-k)^2-m^2]} \nonumber \\
\label{e:anon}
\end{eqnarray}
and the flavor factor is abbreviated as $N_f^{(345k)} = {\rm
tr}_f(T_kT_5T_4T_3)$.
The flavor factor to be associated with the complete $s$-like channel is then
\begin{equation}
f_s^{(345)}= N_f^{(j)}N_f^{(345k)},
\label{e:fl}
\end{equation}
where $N_f^{(j)} = \bar q T_j q$ arises from the choice of incoming quark
and antiquark.    Directly performing the spinor trace leads to the
expression
\begin{eqnarray}
&&A_{\pi\pi\pi\pi}(p_3,p_4,p_5) = -2N_cN_f^{(345k)}[(-k^2p_3p_4 -
kp_4m_\pi^2)I_4(k,p_3+p_4,p_3,0) \nonumber \\
&&+ 2(kp_3)p_4\tilde I_4(k,p_3+p_4,p_3,0) + p_3p_4I_3(p_3+p_4,p_3,0)
\nonumber \\
&&+ (kp_4-k^2)I_3(p_3+p_4,k,0) + (kp_4-p_3p_4-m_\pi^2)I_3(p_3,p_3+p_4,k)
\nonumber \\
&&+ I_2(p_3+p_4,0) + I_2(k,p_3),
\label{e:monster}
\end{eqnarray}
where
\begin{equation}
I_n(a_1,\dots,a_n) = \int\frac{d^4q}{(2\pi)^4} \frac 1{[(q-a_1)^2 - m^2]
\dots[(q-a_n)^2-m^2]}
\label{e:in}
\end{equation}
and $\tilde I_4$ is defined as
\begin{equation}
\tilde I_4(a_1,\dots,a_4)^\mu= \int \frac{d^4q}{(2\pi)^4} \frac {q^\mu}
{[(q-a_1)^2-m^2] \dots[(q-a_4)^2-m^2]}.
\end{equation}
This latter integral, in combination with an external momentum can be
reduced to a linear combination of the $I_n$.   These functions are
listed in Appendix A.

The diagram of Fig.~3 represents one of the six possibilities of ordering
the external momenta.   Another of these is obtained by reversing the arrows
on the internal loop of the vertex, as shown in Fig.~4.    Translation of
this Feynman diagram gives
\begin{eqnarray}
&&A_{\pi\pi\pi\pi}(p_5,p_4,p_3) =\nonumber \\
& & - {\rm Tr}\int\frac{d^4q}{(2\pi)^4} i\gamma_5T_k iS(-q)
i\gamma_5T_3iS(p_3-q)i\gamma_5T_4iS(p_3+p_4-q)  i\gamma_5T_5iS(k-q) .
\nonumber \\
\label{e:arev}
\end{eqnarray}
Some algebra is required to evaluate the traces.    The spinor trace can be
shown to be equivalent to that required for Eq.(\ref{e:anon}).    In addition,
the flavor traces are equal,
\begin{equation}
N_f^{(543k)} = N_f^{(345k)},
\label{e:eqfl}
\end{equation}
so that
\begin{equation}
A_{\pi\pi\pi\pi}(p_5,p_4,p_3) = A_{\pi\pi\pi\pi}(p_3,p_4,p_5).
\label{e:eqs}
\end{equation}
This is a general feature when reversing the loop momenta for the diagrams
required here.    Hence the combination of six terms for the $s$-like
channel reduces to a calculation of 2 times three terms.    The remaining
two ``crossed'' $s$-like graphs for the vertices are shown in Fig.~5.
The associated vertex functions are $A_{\pi\pi\pi\pi}(p_4,p_5,p_3)$
and $A_{\pi\pi\pi\pi}(p_5,p_4,p_3)$, which can be obtained from
Eq.(\ref{e:monster}) on suitably interchanging the arguments.

One may finally construct the scattering amplitude for the $s$-like channel
as
\begin{equation}
-i{\cal M}_\pi^{s{\rm-like}} = -2i[{\cal M}_\pi(p_3,p_4,p_5) +
{\cal M}_\pi(p_4,p_5,p_3) +
{\cal M}_\pi(p_5,p_3,p_4)],
\label{e:mslike}
\end{equation}
where
\begin{equation}
-i{\cal M}_\pi(p_a,p_b,p_c) = \bar v_2i\gamma_5 u_1\delta_{c_1,c_2}\frac{2G}
{1-2G\Pi_{PS}(k^2)}g^3_{\pi q q} N_f^{(j)} A_{\pi\pi\pi\pi} (p_a,p_b,p_c).
\label{e:mgen}
\end{equation}
Since $\Pi_{PS}(k^2)$ and $g_{\pi q q}$ can also be expressed through the
integrals $I_n$,
\begin{eqnarray}
-i\Pi_{PS}(k^2) &=& 4N_c (2I_1(0) - k^2 I_2(0,-k)) \\
g_{\pi q q}^{-2} &=& -4iN_c I_2(0,-k),
\label{e:pig}
\end{eqnarray}
a direct evaluation of this quantity depends only on the evaluation of the
integrals $I_n$, see Appendix A.

For the purpose of obtaining a scattering cross-section, the absolute value
of the scattering amplitude is averaged over initial states and summed over
the final ones.    Since $\sum_{s_1,s_2}|\bar v_2 i\gamma_5 u_1 |^2=2s$, it
follows that
\begin{equation}
\overline{|{\cal M}_\pi(p_a,p_b,p_c)|^2}
= \frac{2G^2 g^6_{\pi q q } (N_f^{(j)})^2 s}{N_c}
\frac{|A_{\pi\pi\pi\pi}(p_a,p_b,p_c)|^2}{|1-2G\Pi_{PS}(k^2)|^2}
\label{e:maverage}
\end{equation}
from which $\overline{|{\cal M}_\pi^{s{\rm-like}}|^2}$ can be constructed.

\subsection{$t$-like channel}

The six possible $t$-like channel graphs are shown in Fig.~6, together with
our labelling of each process.
As an example, we calculate one of these diagrams explicitly.   Considering
the $t-w$ channel depicted in Fig.~6a, one finds
\begin{equation}
-i{\cal M}^{(t,w)}=\bar{v}_2ig_{\pi qq}i\gamma_5 T_5iS(p_1-p_3-p_4)ig_{\pi
qq}
      i\gamma_5 T_4iS(p_1-p_3)ig_{\pi qq}i\gamma_5 T_3 u_1
\delta_{c_1c_2},
\label{tw}
\end{equation}
and therefore
the absolute value of the scattering amplitude averaged over
initial states and summed over the final ones follows as
\begin{eqnarray}
\overline{|{\cal M}^{(t,w)}|^2}&=&\frac{1}{4N_c^2}\sum_{s_1s_2}
                          \frac{-f_{t,w}^2g_{\pi qq}^6
N_c}{(t-m^2)^2(w-m^2)^2}
                          {\rm tr}[v_2\bar{v}_2\gamma_5
(\fsl{p}_1-\fsl{p}_3-\fsl{p}_4+m)\gamma_5\nonumber\\
                    &&\!\!\cdot(\fsl{p}_1-\fsl{p}_3+m)\gamma_5u_1\bar{u}_1
                         \gamma_5(\fsl{p}_1-\fsl{p}_3+m)\gamma_5
                         (\fsl{p}_1-\fsl{p}_3-\fsl{p}_4+m)
\gamma_5].\qquad \nonumber \\
\end{eqnarray}
The trace over flavor matrices has been abbreviated as
$f_{t,w}={\rm tr}(T_5 T_4 T_3)$.
After some tedious algebra for evaluating the spinor trace, one arrives at
the relatively simple form
\begin{eqnarray}
\overline{|{\cal M}^{(t,w)}|^2}&=&\frac{f_{t,w}^2g_{\pi qq}^6}{2N_c}
     \left[\frac{-1}{t-m^2}\left(1+\frac{x-u}{w-m^2}\right)
        +\frac{m_{\pi}^6}{(t-m^2)^2(w-m^2)^2}\right.\nonumber\\
   &&\left.     +\frac{m_{\pi}^2}{(t-m^2)(w-m^2)}\left(2-\frac{y-m^2}{t-m^2}
             -\frac{u-m^2}{w-m^2}\right)\right],
\label{Mtw2}
\end{eqnarray}
in which the invariants $t,u,x,y$ and $w$ occur.
Completely analogous calculations lead to the expressions for the remaining
five channels, $\overline{|{\cal M}^{(u,x)}|^2}$, $\overline{|{\cal
M}^{(v,y)}|^2}$, $\overline{|{\cal M}^{(t,y)}|^2}$,  $\overline{|{\cal
M}^{(u,w)}|^2}$ and  $\overline{|{\cal M}^{(v,x)}|^2}$.    These can be
obtained directly from Eq.(\ref{Mtw2}) by making an appropriate variable
substitution.  To obtain $\overline{|{\cal M}^{(u,x)}|^2}$, one makes the
replacement
\begin{eqnarray}
t&\rightarrow& u\rightarrow v \nonumber \\
w&\rightarrow& x \rightarrow y \rightarrow w
\label{Mux2}
\end{eqnarray}
in Eq.(\ref{Mtw2}), that follows on examining the momentum dependence
of Fig.~6b in comparison with that of Fig.~6a.    Similarly, the
expression for $\overline{|{\cal
M}^{(v,y)}|^2}$ follows from Eq.(\ref{Mtw2}) on substituting
\begin{eqnarray}
t&\rightarrow & v \quad\quad u\rightarrow t \nonumber \\
w&\rightarrow& y \rightarrow x\rightarrow w ,
\label{Mvy2}
\end{eqnarray}
while $\overline{|{\cal M}^{(t,y)}|^2}$ is obtained on writing
\begin{eqnarray}
t &\rightarrow t \quad\quad x\rightarrow x \nonumber \\
u &\rightarrow v \quad\quad w\leftrightarrow y,
\label{Mty2}
\end{eqnarray}
$\overline{|{\cal M}^{(u,w)}|^2}$ from Eq.(\ref{Mtw2}) on writing
\begin{eqnarray}
t&\leftrightarrow & u  \nonumber \\
w&\rightarrow& w \quad\quad x\leftrightarrow y,
\label{Muw2}
\end{eqnarray}
and $\overline{|{\cal M}^{(v,x)}|^2}$ from Eq.(\ref{Mtw2}) on writing
\begin{eqnarray}
t&\rightarrow& v \quad\quad u\rightarrow u \nonumber \\
w&\leftrightarrow& x \quad\quad y\rightarrow y
\label{Mvx2}.
\end{eqnarray}
All expressions have relatively simple analytic forms.
One notices that in the chiral limit, $m_\pi=0$, they
simplify dramatically.

\subsection{{\it st}-like channel}

The $st$-like channel of Fig.~2(c)
is evaluated by separating the vertex for two meson
production from that of single meson production.
 This diagram  is thus subdivided into two pieces, the
first containing the quark-qntiquark scattering amplitude that is mediated
by an exchanged meson connected with a
quark loop as indicated in Fig.~7.    Implicit in this
diagram is that both the clockwise and counterclockwise senses in the
fermion loop are to be evaluated.   Since we will regard pions the final
state, it follows from parity conservation that the intermediate mesonic
state must be a scalar.   In this case, this is then taken to be the
$\sigma$ meson. Since the two flavor NJL model is
considered here, the two pions that emerge from this vertex must be charge
neutral.   Translating Fig.~7, one has
\begin{eqnarray}
A_{\sigma \pi\pi}(p) &=& iD_S(p^2)(ig_{\pi q q})^2(-2N_c)
{\rm tr}_f{T_3T_4}\nonumber \\
&\times&\int\frac{d^4q}{(2\pi)^4}
{\rm tr}_\gamma[i\gamma_5iS(q)i\gamma_5iS(q+p_3)iS(q-p_4)].
\nonumber \\
\label{e:v2p}
\end{eqnarray}
The factor 2 arises from considering both senses of the fermion loop,
 and again ${\rm tr}_\gamma$ is the
trace on Dirac indices.   In this expression, $D_S(p^2)$ is the
quark-quark scattering amplitude in the scalar $\sigma$ channel, i.e.
\begin{equation}
D_S(p^2) = \frac{2G}{1-2G\Pi_S(p^2)}.
\label{e:ds}
\end{equation}
In principle, the exchange of the $\sigma$ meson within the model
as required here, involves a complex function, since the $\sigma$
is a resonance.   It is, however, only weakly unbound, with
$m_\sigma^2\simeq 4m^2 + m_\pi^2$, and consequently here only
principle values of the integrals $I_n$ entering into the evaluation
of the irreducible polarization $\Pi_S(p^2)$ is used.

By making the substitution $q\rightarrow -q$ in the second integral 
of Eq.(\ref{e:v2p}) and
performing the trace, one arrives at the following expression for this
function,
\begin{equation}
A_{\sigma\pi\pi}(p) = 16 mN_c g_{\pi q q}^2 D_S(p^2)[I_2(p_3,-p_4) + p_3
\cdot p_4I_3(0,
p_3,-p_4)].
\label{e:v2calc}
\end{equation}

This graph is then embedded into the hadronization graph of Fig.~2(c).
Attaching the vertex $A_{\sigma\pi\pi}$ to both the incoming quark line, as
was drawn originally in Fig.~2(c)
plus the incoming antiquark line, which is also a required configuration,
leads to two terms,
\begin{eqnarray}
{\cal M}^{(st)} &=& \bar v(p_2)[i\gamma^5 T ig_{\pi q q}\frac
{i(\not\! p_5 - \not \! p_2 + m)}{(p_5-p_2)^2 -m^2}A_{\sigma\pi\pi}(p_3+p_4)
\nonumber \\
&+& A_{\sigma\pi\pi}(p_3+p_4) \frac{i(\not \! p_1 - \not \! p_5 + m)}{
(p_1-p_5)^2 + m^2} i\gamma^5 T ig_{\pi q q}] u(p_1) \nonumber \\
&=& {\cal M}_1^{(st)} + {\cal M}_2^{(st)} .
\label{e:missing}
\end{eqnarray}
The averaged transition matrix element then follows as
\begin{eqnarray}
\overline{|{\cal M}^{(st)}|^2} &=& \frac 1{2N_n} g_{\pi q
q}^2|A_{\sigma\pi\pi}(p_3+
p_4)|^2\left \{\frac 1{w-m^2} + \frac 1{v-m^2}\right \}^2 \nonumber\\
&\times& \{m_\pi^2(4m^2 -s) + (v-m^2-m_\pi^2)(w-m^2-m_\pi^2)\}.
\label{e:newm}
\end{eqnarray}
   Since $p_5$ is
the momentum of the pion line that is attached singly, we use the notation
${\cal M}^{(st)}(p_5)$ to denote this fact.

\subsection{Full scattering amplitude}

The full scattering amplitude is comprised of all $s$-like, $t$-like
and $st$-like contributions,
{\it i.e.}
\begin{eqnarray}
{\cal M}&=&2{\cal M}_{\pi}(p_3,p_4,p_5)+2{\cal M}_{\pi}(p_4,p_5,p_3)
         +2{\cal M}_{\pi}(p_5,p_3,p_4)\nonumber\\
&&{}     +{\cal M}^{(t,w)}+{\cal M}^{(u,x)}+{\cal M}^{(v,y)}
         +{\cal M}^{(t,y)}+{\cal M}^{(u,w)}+{\cal M}^{(v,x)} \nonumber \\
&&{} + {\cal M}^{(st)}(p_3) + {\cal M}^{(st)}(p_4) + {\cal M}^{(st)}(p_5).
\label{mtotal}
\end{eqnarray}
As such, the cross-section requires not only the averaged terms that were
given in Eqs.(\ref{e:maverage}) and (\ref{Mtw2}) - (\ref{Mvx2}), but also mixed
terms
that arise on building the modulus.   There are many of these and we have
listed some of them  in Appendix B.

\section{The scattering cross-section for $
\lowercase{q\bar q}\rightarrow \pi\pi\pi$
in terms of extended Mandelstam-like variables}

\subsection{Cross section}

The notion of relativistic invariance is an important calculational guide in
evaluating matrix elements.   As such, it is particularly useful to express
the integrated cross section in terms of relativistic invariants only.   To
do this in the
case when two particles are incident and two are exiting is a standard
textbook exercise, see for example \cite{itz}.   However, when three particles
are
involved in the exit channel, as is the case here, this becomes somewhat
more complex, and we are required to derive a formal expression.   The only
restriction that occurs in our derivation is that the incident particles
with momenta $p_1$ and $p_2$ refer to particles with the same mass, {\it
i.e.}
$p_1^2=p_2^2=m^2$, while the exiting particles with momenta $p_3$, $p_4$,
and $p_5$ also are bound by the constraint $p_3^2=p_4^2=p_5^2=m_\pi^2$.
The generalization to arbitrary mass particles in describing the reaction
of $a+b\rightarrow c+d+e$ is obvious.

Starting from the definition of the cross section
\begin{eqnarray}
\sigma&=&\int\frac{(2\pi)^4 \delta^{(4)}(p_3+p_4+p_5-p_1-p_2)\,
         \overline{|{\cal M}(p_1,p_2,p_3,p_4,p_5)
|^2}}{|\vec{v}_{rel}|2E_12E_2}
\nonumber\\
      & &\times\frac{d^3p_3}{(2\pi)^3 2E_3}
             \frac{d^3p_4}{(2\pi)^3 2E_4}\frac{d^3p_5}{(2\pi)^3 2E_5}
\label{sig0}
\end{eqnarray}
and the definition of the extended Mandelstam-like variables of
Eq.(\ref{e:exmand}),
we
use the identity
\begin{equation}
\int dp^0_5\,\Theta(p^0_5)\,\delta(p^2_5-m_{\pi}^2)\,2p^0_5=1
\end{equation}
and perform the $p_5$ integration, giving
\begin{eqnarray}
\sigma&=&\int\frac{\overline{|{\cal M}(s,t,u,v,w,x,y)|^2}}{(2\pi)^5
\,|\vec{v}_{rel}|2E_12E_2}
       \,\frac{d^3p_3}{ 2E_3}\,\frac{d^3p_4}{ 2E_4}
       \,\delta((p_1+p_2-p_3-p_4)^2-m_{\pi}^2) \nonumber \\
   &\times&    \,\Theta(p^0_1+p^0_2-p^0_3-p^0_4).
\label{e:sig1}
\end{eqnarray}
{}From energy conservation, the argument of the $\Theta$ function can be
evaluated to be $p_1^0+p_2^0-p_3^0-p_4^0=E_1+E_2-E_3-E_4=E_5>0$, so that
the $\Theta$ function takes the value 1.
The argument of the $\delta$ function can be reexpressed in terms of the
extended Mandelstam-like variables either as
\begin{equation}
(p_1+p_2-p_3-p_4)^2 -m_\pi^2 = s + w + x + y - 3m^2-3m_\pi^2
\label{e:man1}
\end{equation}
or as
\begin{equation}
(p_1+p_2-p_3-p_4)^2 -m_\pi^2 = s+t+u+v-3m^2-3m_\pi^2,
\label{e:man2}
\end{equation}
and the fact that the $\delta$ function contributes only when the right hand
side of Eqs.(\ref{e:man1}) or (\ref{e:man2}) is zero leads to the energy
momentum relations already stated in Eqs.(\ref{e:newrel1}) or
(\ref{e:newrel2}).

    As a consequence of their
definitions in Eq.(\ref{e:exmand}), the variables $s$, $t$, $u$ and $w$
are unaffected by the $p_5$ integration.   On the other hand, $v$, $x$ and
$y$ are altered,
\begin{eqnarray}
v=(p_1-p_5)^2\;\;\;\;\;\;\;\;&\to& (p_2-p_3-p_4)^2 \nonumber\\
x=(p_1-p_4-p_5)^2 &\to& (p_2-p_3)^2 \nonumber\\
y=(p_1-p_3-p_5)^2 &\to& (p_2-p_4)^2 .
\end{eqnarray}
The next task lies in transforming the element of integration and
determining the appropriate Jacobian.   To this end, we
 denote $\theta_{3,4}$ as being the angle between the vectors
$\vec p_1$ and $\vec p_3$ or $\vec p_4$.   Since we work in the c.m.
reference frame, $\vec p_1$ and $\vec p_2$ lie in opposite directions so
that
the angle between $\vec p_2$ and $\vec p_3$ or $\vec p_4$
is $(\pi-\theta_{3,4})$.   In
addition, $|\vec p_1|=|\vec p_2|$,
 so that the
extended Mandelstam-like variables of Eq.(\ref{e:exmand}) take on the forms
\begin{mathletters}
\begin{eqnarray}
t
 &=&m^2+m_{\pi}^2-2E_1E_3+2|\vec{p}_1||\vec{p}_3|
                                 \cos \theta_3\label{t1}\\
u
 &=&m^2+m_{\pi}^2-2E_1E_4+2|\vec{p}_1||\vec{p}_4|
                                 \cos \theta_4\label{u1}\\
v
 &=&m^2+2m_{\pi}^2-2E_1E_3
    -2E_1E_4
+2E_3E_4\nonumber \\
 && -2|\vec{p}_1||\vec{p}_3|\cos \theta_3
    -2|\vec{p}_1||\vec{p}_4|\cos \theta_4    \nonumber\\
 && -2|\vec{p}_3||\vec{p}_4|
        (\sin \theta_3\sin \theta_4\cos(\varphi_4-\varphi_3)
         +\cos \theta_3\cos \theta_4)\label{v1}\\
w
 &=&m^2+2m_{\pi}^2-2E_1E_3
    -2E_1E_4
+2E_3E_4\nonumber\\
 && +2|\vec{p}_1||\vec{p}_3|\cos \theta_3
    +2|\vec{p}_1||\vec{p}_4|\cos \theta_4\nonumber\\
 && -2|\vec{p}_3||\vec{p}_4|
     (\sin \theta_3\sin \theta_4\cos(\varphi_4-\varphi_3)
         +\cos \theta_3\cos \theta_4)\label{w1}\\
x
 &=&m^2+m_{\pi}^2-2E_1E_3-2|\vec{p}_1||\vec{p}_3|
                                 \cos \theta_3\label{x1}\\
y
 &=&m^2+m_{\pi}^2-2E_1E_4-2|\vec{p}_1||\vec{p}_4|
                                 \cos \theta_4.
\label{tuvwxy}
\end{eqnarray}
\end{mathletters}
For conciseness, we have used the abbreviation $E^2_{3,4} = |\vec p_{3,4}|
^2 + m_\pi^2$, but draw the attention of the reader to the fact that we
consider $|\vec p_{3,4}|$ together with the angles as variables, and not
$E_{3,4}$, althougth this would be equivalent.    By contrast, $E_1$ and
$E_2$ are fixed by the centre of mass condition, $E_1 = E_2 =\sqrt s/2$.
    Conversely the inverse
transformation can be obtained.     By adding and subtracting Eq.(\ref{t1})
from Eq.(\ref{x1}), one can easily find the determining equations for
$|\vec p_3|$ and $\theta_3$.    A similar manipulation of Eqs.(\ref{u1})
and (\ref{tuvwxy}) yields those for $|\vec p_4|$ and $\theta_4$.    Having
done this, Eqs.(\ref{v1}) or (\ref{w1}) can be easily solved for
$\cos(\varphi_4-\varphi_3)$.  The exact form of this transformation is
\begin{mathletters}
\begin{eqnarray}
t-x=4|\vec{p}_1||\vec{p}_3|\cos \theta_3 \;
&\Rightarrow&\;  |\vec{p}_3|\cos
\theta_3=\frac{t-x}{4|\vec{p}_1|}, \label{g1} \\
u-y=4|\vec{p}_1||\vec{p}_4|\cos \theta_4\;
&\Rightarrow&\; |\vec{p}_4|\cos
\theta_4=\frac{u-y}{4|\vec{p}_1|},\label{g2}\\
t+x=2m^2+2m_{\pi}^2-4E_1E_3\;
&\Rightarrow&\; E_3=\frac{2m^2+2m_{\pi}^2-t-x}{4E_1},\label{g3}\\
u+y=2m^2+2m_{\pi}^2-4E_1E_4\;
&\Rightarrow&\; E_4=\frac{2m^2+2m_{\pi}^2-u-y}{4E_1},
\label{g4}
\end{eqnarray}
\end{mathletters}
while the  expression for $\cos \varphi_{34}$ is
\begin{equation}
\cos \varphi_{34}=\frac{t+u-w-m^2+2E_3E_4
                        -2|\vec{p}_3||\vec{p}_4|\cos \theta_3\cos \theta_4 }
                       {2|\vec{p}_3||\vec{p}_4|\sin \theta_3\sin \theta_4}.
\label{cosphi34}
\end{equation}

Returning to the   integrand in Eq.(\ref{e:sig1}), one sees that it
 depends only on the extended
Mandelstam-like variables, that in turn can be written in terms of $|\vec
p_3|$,
$|\vec p_4|$, $\theta_3$, $\theta_4$ and the difference
$\varphi_{34}=\varphi_4-\varphi_3$.
Thus
\begin{equation}
\int d\varphi_3 d\varphi_4 = 2\pi\int d\varphi_{34} .
\label{e:phis}
\end{equation}
There are five independent integration variables,
$|\vec p_3|$, $|\vec p_4|$, $\theta_3$, $\theta_4$ and $\varphi_{34}$.
On the other hand, there are seven extended Mandelstam variables.
    As can be seen from
Eqs.(\ref{v1}) and (\ref{w1}), both $v$ and $w$ are functions of the variable
$\varphi_{34}$;   we arbitrarily choose to discard $v$.   In keeping with
convention, the other unrelated variable is $s$, the square of the center
of mass energy of the incoming particles.    We thus are required to
relate the multidimensional volume element in the variables
$|\vec p_3|$, $|\vec p_4|$, $\theta_3$, $\theta_4$ and $\varphi_{34}$ to
the extended set $t$, $u$, $w$, $x$ and $y$, which follows
as
\begin{equation}
2|J|\, d|\vec p_3|\, d|\vec p_4|\, d\theta_3 \, d\theta_4\, d\varphi_{34}
= dt\; du\; dw\; dx\; dy ,
\label{e:trans}
\end{equation}
where the factor 2 has been introduced together with the reduction of the
integration region of $\varphi_{34}$ from $[0,2\pi]$ to $[0,\pi]$.   After
some calculation, one finds the Jacobian of the transformation to be
\begin{equation}
| J|=128\, |\vec{p}_1|^2\,|\vec{p}_3|^3\,|\vec{p}_4|^3\,
            \sin^2\theta_3 \, \sin^2\theta_4 \, \frac{E_1^2}{E_3E_4}\,
            |\sin \varphi_{34}|.
\label{e:jake}
\end{equation}
Using the information from Eqs.(\ref{e:man1}) and (\ref{e:jake}), the
cross section takes the form
\begin{eqnarray}
\sigma
  &=&\int\frac{\overline{|{\cal M'}(s,t,u,w,x,y)|^2}}
               {(2\pi)^4\,4|\vec{p}_1|\sqrt{s}}
         \,\delta(s+w+x+y-3m^2-3m_{\pi}^2)\nonumber\\
   &&\times\frac{|\vec{p}_3|^2\,\sin \theta_3\,|\vec{p}_4|^2\,\sin
\theta_4\,
               E_3\,E_4\,dt\,du\,dw\,dx\,dy}
              {2E_3\, 2E_4\,256 |\vec{p}_1|^2|\vec{p}_3|^3|\vec{p}_4|^3
               \sin^2\theta_3 \, \sin^2\theta_4\,E_1^2\,|\sin
\varphi_{34}|},
\\
\label{e:interm}
\end{eqnarray}
which can be written in a simpler form as
\begin{eqnarray}
\sigma(s) &=& \frac 1{64(2\pi)^4}
\int\, dt\, du\, dw \,dx\,dy \frac 1{N(s(s-4m^2))^{3/2}}
\overline{|{\cal M}'(s,t,u,w,x,y)|^2}  \nonumber \\
& & \times\quad\quad\delta(s+w+x+y-3m^2-3m_\pi^2),
\label{e:pretty}
\end{eqnarray}
where use has been made of the relationship $4|\vec p_1|^2 =
s-4m^2$.   In both Eqs.(\ref{e:interm}) and (\ref{e:pretty}), the notation
$M'$ has been introduced to remind us that the variable $v$ in the
calculated matrix elements is to be replaced by $(-s-t-u+3m^2+3m_\pi^2)$
and the factor
\begin{equation}
N=2|\vec p_3||\vec p_4| \sin\theta_3\sin\theta_4\sqrt{1-\cos^2\varphi_{34}}
\label{e:n}
\end{equation}
has been abbreviated in the denominator.   Clearly $N$ must also be
expressed in terms of the extended Mandelstam-like invariants.  One finds
 the rather unwieldy expression
\begin{eqnarray}
N &=& N(s,t,u,w,x,y) \nonumber \\
&=&\left[ \frac{-1}{4s(s-4m^2)}[(2m^2+2m_{\pi}^2)(t-x-u+y)+2ux-2ty]^2
                                                          \right.\nonumber\\
 &&   +m_{\pi}^2\left(\frac{(t-x)^2+(u-y)^2}{s-4m^2}
-\frac{(2m^2+2m_{\pi}^2-t-x)^2+(2m^2+2m_{\pi}^2-u-y)^2}{s}
                \right)\nonumber\\
 &&   -(t+u-w-m^2)^2 + 4m_{\pi}^4\nonumber\\
 &&   +(t+u-w-m^2)\left(
\frac{(2m^2+2m_{\pi}^2-t-x)(2m^2+2m_{\pi}^2-u-y)}{s}
                                                     \right.\nonumber\\
 &&\left.\left.        -\frac{(t-x)(u-y)}{s-4m^2} \right)\right]^{1/2}.
\label{e:bign}
\end{eqnarray}
In the chiral limit, this function simplifies somewhat.

As a final step, the integration on $w$ can be performed and we obtain
our final expression for the cross section:
\begin{equation}
\sigma(s) = \frac 1{64(2\pi)^2}\int dt\, du\,dx\,dy\,\frac
{ \overline {|{\cal M}''(s,t,u,x,y)|^2} }
{(s(s-4m^2))^{3/2} N(s,t,u,x,y)} ,
\label{e:lastsig}
\end{equation}
which involves a fourfold integration.   The notation $M''$ serves here to
remind one that $w$ has been integrated out, and is to be replaced by
$(3m^2+3m_\pi^2-s-x-y)$ in these matrix elements.   Using an obvious
notation, the function
\begin{equation}
N(s,t,u,x,y) = N(s,t,u,w=3m^2+3m_\pi^2-s-x-y,x,y)
\label{e:nagain}
\end{equation}
can be obtained from Eq.(\ref{e:bign}).   This result generalizes the
well-known form for the cross section for two incoming particles to two
outgoing particles,
\begin{equation}
\sigma(s) = \int\frac { \overline{|{\cal M}'(s,t)|^2} }
{16\pi s(s-4m^2)}  dt
\label{e:sig2}
\end{equation}
in which ${\cal M}'(s,t)$ denotes matrix elements ${\cal M}'(s,t) =
{\cal M}(s,t,u=2m^2+2m_\pi^2-s-t)$, for $q\bar q\rightarrow \pi\pi$ say.

\subsection{Limits}

In reactions of the type $a+b\rightarrow c+d$, for which the cross section
of the form Eq.(\ref{e:sig2}) holds, the determination of the limits of
integration of the single variable $t$ is simple:  it is always
related to one angular variable by virtue of its definition, and the
minimum and maximum values of this angular
variable lead to a minimum and maximum
value of $t$ respectively.    For the case that is implicit in
Eq.(\ref{e:sig2}), a simple analytic form for $t_{max}$ and $t_{min}$ can be
found.   One has
\begin{equation}
t_{max}=m^2+m_{\pi}^2-\frac{s}{2}+\sqrt{s-4m^2}\sqrt{\frac{s}{4}-m_{\pi}^2}
\end{equation}
and
\begin{equation}
t_{min}=m^2+m_{\pi}^2-\frac{s}{2}-\sqrt{s-4m^2}\sqrt{\frac{s}{4}-m_{\pi}^2}
\end{equation}
for the case of different particles and
\begin{equation}
t_{min}=m^2+m_{\pi}^2-\frac{s}{2}
\end{equation}
when there are identical particles in the final state.

By contrast, in our case where there are three outgoing particles in the
final state, we have four integration variables $t$, $u$, $x$ and $y$, whose
ranges have to be determined, and which are laid fixed by the maximum and
minimum values that these variables can take according to Eqs.
(\ref{t1})-(\ref{tuvwxy})
which are in turn set by the
 appropriate minimal and maximal values of the angular and
energy variables that occur therein.    For this reason, it is difficult to
give simple expressions for these limits and we thus start by giving
their ranges.    Obvious constraints on the particle energies
and angles involved are
\begin{mathletters}
\begin{eqnarray}
E_3 &\ge& m_{\pi} \label{1}\\
E_4 &\ge& m_{\pi} \label{2}\\
E_5 &\ge& m_{\pi} \label{3}\\
\cos^2 \Theta_3 &\le& 1 \label{4}\\
\cos^2 \Theta_4 &\le& 1 \label{5}\\
\cos^2 \varphi_{34} &\le& 1 \label{6}.
\end{eqnarray}
\end{mathletters}
These constraints on the angles must be translated into conditions for
$t$, $u$, $x$, and $y$,
while an expression for $\cos \varphi_{34}$ can be obtained on inverting
Eq.(\ref{w1}) for $w$.
Now,   introducing the abbreviations
\begin{mathletters}
\begin{eqnarray}
\tilde{t}&=&t-m^2-m_{\pi}^2 \label{m1} \\
\tilde{u}&=&u-m^2-m_{\pi}^2 \label{m2} \\
\tilde{x}&=&x-m^2-m_{\pi}^2 \label{m3} \\
\tilde{y}&=&y-m^2-m_{\pi}^2 \label{m4}
\end{eqnarray}
\end{mathletters}
we can reexpress Eqs.(\ref{1}) and (\ref{2}) as
\begin{equation}
-\tilde{t}-\tilde{x}-2m_\pi\sqrt{s}\ge 0
\label{1a}
\end{equation}
\begin{equation}
-\tilde{u}-\tilde{y}-2m_\pi\sqrt{s}\ge 0,
\label{2a}
\end{equation}
which follows on using the equations (\ref{g3}) and (\ref{g4}).
The equation (\ref{3}), on the other hand, becomes
\begin{equation}
\sqrt{s}-E_3-E_4-m_{\pi}\ge 0,
\label{e:cond5}
\end{equation}
on using the conservation of energy condition $E_3+E_4+E_5=\sqrt{s}$ in the
c.m. system.      $E_3$ and $E_4$ can be eliminated from Eq.(\ref{e:cond5})
using (\ref{g3}) and (\ref{g4}).    Then it reads
\begin{equation}
\sqrt{s}+\frac{\tilde{t}+\tilde{x}}{2\sqrt{s}}
    +\frac{\tilde{u}+\tilde{y}}{2\sqrt{s}}-m_{\pi}\ge 0.
\label{e:bound}
\end{equation}
This expression can be further utilized to set an upper bound on either
$\tilde{t}+\tilde{x}$ or $\tilde{u}+\tilde{y}$.    Let us regard that for
$\tilde{t}+\tilde{x}$.    Eliminating $\tilde{u}+\tilde{y}$ using
Eq.(\ref{2a}), Eq.(\ref{e:bound}) leads to the bound
\begin{equation}
\tilde{t}+\tilde{x}\ge -2s+4\sqrt{s}m_{\pi} \label{3a}.
\label{e:boundxt}
\end{equation}

We next examine the angular condition Eq.(\ref{4}).   In terms of $t$ and
$x$, or $\tilde{t}$ and $\tilde{x}$, this reads
\begin{equation}
\left ( \frac{\tilde{t}-\tilde{x}}{4|\vec{p}_1||\vec{p}_3|}\right)^2 \le1,
\label{e:angcond1}
\end{equation}
where Eq.(\ref{g1}) was used.
By eliminating $4|\vec p_1|^2 = s-4m^2$ and $|\vec p_3|^2 = E_3^2
-m_\pi^2$ with the aid of Eq.(\ref{g3}), this condition becomes
\begin{equation}
(\tilde{t}-\tilde{x})^2 s \le (s-4m^2)((\tilde{t}+\tilde{x})^2-4sm_{\pi}^2).
\label{4a}
\end{equation}
In an analogous fashion, the angular condition Eq.(\ref{5}) can be
rearranged to give a condition on $\tilde{u}-\tilde{y}$:
\begin{equation}
(\tilde{u}-\tilde{y})^2 s \le(s-4m^2)((\tilde{u}+\tilde{y})^2-4sm_{\pi}^2).
\label{5a}
\end{equation}
The final angular constraint, Eq.(\ref{6}) is rewritten in terms of the
Mandelstam-like variables on considering Eq.(\ref{cosphi34}), replacing
$w$ by $3m^2+3m_\pi^2-s-x-y$ and the appropriate energies from
Eqs.(\ref{g3}) and (\ref{g4}).    This then takes the form
\begin{equation}
\frac{\left ( s+\tilde{t}+\tilde{u}+\tilde{x}+\tilde{y}+m_{\pi}^2
               +\alpha\gamma/s -\beta\delta/\tilde{s}          \right )^2}
     {4(E_3^2-m_{\pi}^2)(E_4^2-m_{\pi}^2)
      -2\delta(E_3^2-m_{\pi}^2)/\tilde{s}
      -2\beta (E_4^2-m_{\pi}^2)/\tilde{s}
      +\beta^2\delta^2/\tilde{s}^2        }   \le 1,
\label{6a}
\end{equation}
where the abbreviations
\begin{eqnarray}
\alpha=\frac{\tilde{t}+\tilde{x}}{\sqrt{2}}, \;\;&\;\;\;\;\;&\;\;
\beta =\frac{\tilde{t}-\tilde{x}}{\sqrt{2}}, \label{alphabeta}\\
\gamma=\frac{\tilde{u}+\tilde{y}}{\sqrt{2}}, \;\;&\;\;\;\;\;&\;\;
\delta=\frac{\tilde{u}-\tilde{y}}{\sqrt{2}}  \label{gammadelta}
\end{eqnarray}
have been introduced and $\tilde{s} = s-4m^2$                  .

The integration region for the variables $t$, $u$ $x$ and $y$ must be
determined from the relations (\ref{1a}), (\ref{2a}), (\ref{e:boundxt}),
(\ref{4a}), (\ref{5a}) and (\ref{6a}).
{}From (\ref{1a}), it follows immediately that
\begin{equation}
\alpha\le -\sqrt{2s} m_\pi,
\label{1b}
\end{equation}
while Eq.(\ref{4a}), considering the equal to sign, describes a hyperbola
in the variables $\alpha$ and $\beta$
with defining equation
\begin{equation}
\frac{\alpha^2}{2sm_{\pi}^2}-\frac{\beta^2}{2\tilde{s}m_{\pi}^2}=1.
\label{e:hyper}
\end{equation}
The hyperbola has semi-axes $a=\sqrt {2sm_\pi^2}$ and $b=\sqrt{2\tilde{s}
m_\pi^2}$, and asymptotes $\beta = \pm b\alpha/a$.
 In view of Eq.(\ref{1b}), only the negative hyperbola must be
considered, which
 is depicted in Fig.~8.     In order to determine
whether the integration region lies within or without this function, it
suffices to choose a point in the plane.   Choosing $\beta=0$ and
$\alpha<-a$ leads to the relations
\begin{eqnarray}
E_3 &=& \frac{-\tilde{t}-\tilde{x}}{2\sqrt{s}} = - \frac{\alpha}{\sqrt{2s}}
\\
2(\tilde{t} + \sqrt{s} E_3) &=& \sqrt 2\beta = 0
\label{e:interm2}
\end{eqnarray}
so that
\begin{equation}
(s-4m^2)(E_3^2-m_{\pi}^2)-(\tilde{t}+\sqrt{s}E_3)^2>0,
\end{equation}
{\it i.e.} the condition Eq.(\ref{4a}) is fulfilled.    The integration region
thus lies within the hyperbola.
The bound on $\tilde{t}+ \tilde{x}$ that was found in Eq.(\ref{e:boundxt})
can be translated to the variable $\alpha$ and gives rise to the vertical
 line
\begin{equation}
\alpha\ge -\sqrt2  s + 4\sqrt{\frac s 2}m_\pi
\label{e:vertbound}
\end{equation}
shown in Fig.~7 with dots and which cuts off the integration regime
to the left.  This defines a clearly delineated
 area $F_1$ in the $\alpha-\beta$ plane for the integration region.
Note that
one could transform the conditions on $\alpha$ and $\beta$ back to conditions
on $\tilde{t}$ and $\tilde{x}$.    This results in a hyperbola for these
functions that is obtained on rotation of the $\alpha$ and $\beta $ axes by
$\pi/4$.     It is not necessary to find this form explicitly, because the
checks that are done to evaluate the integral will determine the range of
integration numerically using the conditions on $\alpha$ and $\beta$.

In an analogous fashion, it is possible to construct an area $F_2$ in
the variables $\gamma-\delta$   that are related to $\tilde{u}$ and
$\tilde {y}$ that is delineated by the analogous condition
\begin{equation}
\gamma\ge-\sqrt{2}s+4\sqrt{\frac{s}{2}}m_{\pi}
\label{reGerade2}
\end{equation}
plus a hyperbola in the $\gamma-\delta$ plane.     Taken together, these
two areas $F_1\times F_2$ span a superset of  points in $t$-$u$-$x$-$y$
 space that constitute the region of integration, subject to the condition
Eq.(\ref{6a}) being fulfilled.

\section{Numerical Calculations}

In this section, we examine the  reaction $u\bar u \rightarrow
\pi^+\pi^-\pi^0$.    This particular process has been selected, because
the analogous two particle final state hadronization process
$u\bar u\rightarrow \pi^+\pi^-$ is the dominant channel \cite{pr} for
$u\bar u\rightarrow MM'$, where $M$, $M'$ are mesonic members of the
SU$_f$(3) multiplet.    In order to make a sensible comparison, we
calculate the transition rate for this process, which is given in
terms of the cross section $\sigma(s)$ as
\begin{equation}
\omega(s) = |\vec v_{rel}|\sigma(s),
\label{e:omega}
\end{equation}
where $\vec v_{rel}$ is the relative velocity of the incoming quarks,
and which is given as
\begin{equation}
|\vec v_{rel}| = 2\sqrt{1-\frac{4m^2}s}
\label{e:relvel}
\end{equation}
in the centre of mass frame.   Our numerical calculations are based on
a Monte Carlo procedure for evaluating the fourfold integral in the
expression Eq.(\ref{e:lastsig}) for the cross section.
In implementing this, it proves useful to alter the integration region.
As already mentioned, this consists of surfaces $F_1$ and $F_2$ in the
$\alpha$-$\beta$ and $\gamma$-$\delta$ plane respectively and is bounded by
the further condition Eq.(\ref{6a}).    First we extend the surfaces
$F_1$ and $F_2$ to the
 enlarged regions $W_1$ and $W_2$ that contain them, but
which have simpler boundaries.    The triangular region bounded by the
asymptotes of the hyperbolae together with the vertical restriction on the
value of the ordinate as indicated in Fig.~8 for the $\beta$-$\alpha$ plane
is chosen as the extension of $F_1$ to $W_1$, and a similar construction is
made for $W_2$ in the $\gamma$-$\delta$ plane.   This has the advantage
that random points can be chosen to be equally distributed in this
triangular area for the Monte Carlo procedure.   On actually performing the
numerical integration, the caveat is imposed that
the function to be integrated takes
the required functional value in $F_{1}$, $F_2$
 but which is zero otherwise, together with the restriction
Eq.(\ref{6a}).
The delineation of the integration regions
for $\alpha$ and $\beta$ is
given via
\begin{eqnarray}
-\alpha_{max} &<& \alpha < 0 \nonumber \\
-\beta_{max} &<& \beta < \beta_{max}
\label{e:newdelin}
\end{eqnarray}
with $\alpha_{max}=2/\sqrt2 - \sqrt s/\sqrt 2 m_\pi$ and $\beta_{max} =
\alpha_{max}\sqrt{\tilde{s}}/\sqrt s$.

In order to evaluate the transition rate, the individual contributions
to ${\cal M}$ that occur in Eq.(\ref{mtotal}) must be specified.
These are in turn dictated by the traces of the flavor factors, which
sometimes vanish.   For the specific combination of $\pi^+\pi^-\pi^0$ in
the final state, the $s$-like channel contribution contains all six
possible graphs,
\begin{equation}
{\cal M}^{s{\rm -like}} = 2{\cal M}_\pi(p_3,p_4,p_5) + 2{\cal M}_\pi
(p_4,p_5,p_3) + 2{\cal M}_\pi(p_5,p_3,p_4),
\label{e:mscontribs}
\end{equation}
while in the $t$-like channel, only three possible graphs contribute,
\begin{equation}
{\cal M}^{t{\rm-like}} = {\cal M}^{(t,w)} + {\cal M}^{(v,y)} + {\cal
M}^{(t,y)}.
\label{e:mtcontribs}
\end{equation}
In the $st$-like channel, there are only four possibilities.   Here the
single outgoing meson must be a $\pi^0$, while the loop contains the
$\pi^+$ and $\pi^-$ outgoing in the final state.   The two loop directions
plus the attachment to either the incoming quark or antiquark line
comprise these possibilites.   They are fully summarized in the function
\begin{equation}
{\cal M}^{st{\rm-like}} = {\cal M}^{(st)}(p_5).
\label{e:stcombs}
\end{equation}
Rather than evaluate the complete transition rate
that is constructed from ${\cal M}
 ={\cal M}^{s{\rm -like}} + {\cal M}^{t{\rm-like}}
+ {\cal M}^{st{\rm-like}}$, we evaluate first the
transition rate in each particular channel, in order to illustrate the
magnitude that each type of channel contributes.
We comment briefly on our numerical procedure.  In the $s$-like channel,
the pseudoscalar polarization function $\Pi_{PS}(k^2)$ is required from 
Eq.(\ref{e:mpi}), while for the $st$-like channel, the scalar polarization
function $\Pi_S(k^2)$ enters as in Eq.(\ref{e:v2calc}), corresponding to the
exchange of $\pi$ or $\sigma$ mesons respectively.    At zero temperature,
the case studied here, one has that at $k^2=m_\pi^2$, the $\pi$ is strongly
bound, while at $k^2=m_\sigma^2$, the $\sigma$ is weakly unbound, since
$m_\sigma^2\simeq4m^2+m_\pi^2$.   Thus the $\sigma$ already has a small 
width, representing the unphysical decay into two quarks.   At higher values
of $k^2$, this unphysical width grows, and at large values of $k^2$, the
pion can also develop a width.     We have chosen to ignore such 
unphysical decays into quark constituents.
   This is done primarily for reasons of numerical
simplicity -- only the principle values of the integrals $I_n$ that enter
into
$\Pi_{PS}(k^2)$ and  $\Pi_S(k^2)$ are considered --  and this allows one
to focus on the role of the triangular vertex graphs, which are the 
physical cause of
the decays into the mesons.      This procedure simply assumes that the
intermediate mesonic states are stable with regard to decay into their
quark constituents, which is not unreasonable.     Note that the imaginary
parts of the $I_n$ that enter into the irreducible polarization functions
commence always at the threshold $k^2=4m^2$.    A detailed discussion of 
these functions giving their analytic structure,
 together with graphic illustrations as a function of
$k_0^2$ and $\vec k^2$
can be found in Ref.\cite{rehkl}.

For our numerical results, the parameters $\Lambda=851$ MeV, $G\Lambda^2
= 2.87$, and $m_0=5.2$ MeV have been used.   These lead to the values
$f_\pi=93$ MeV, $m_\pi = 135$ MeV and $\langle\bar\psi\psi\rangle = (-250$
MeV$)^3$.   Calculated values of the quark mass yield $m=265$ MeV and the
pion quark coupling as being $g_{\pi q q} = 2.85$.

  Figures 9,10, and 11
display the results for the $s$-like, $t$-like, and $st$-like channels
individually.   One sees that the $st$-like channel provides the largest
yield, being substantially larger than the $s$-like channel (an order of
magnitude) and the $t$-like channel (two orders of magnitude).
Thus, one sees that within the set of graphs that are of the same order
in the $1/N_c$ expansion in fact have a very different relevance.  It is
important to try to understand this, and we can do so on a heuristic
level: one notes that both the $s$ and $st$
like channels contain a quark loop.    The fewer the number of quark lines
occurring in this loop, the more divergent, or less convergent this diagram
will be.   Graphs
such as that occurring for  the $s$-like channel contain a quark loop with
 four internal
quark lines in contrast  to the graphs of the $st$-like channel, which has
a quark loop with three internal quark lines.   One may thus expect that the
$st$-like channel will dominate over the $s$ channel.    Our findings and
this
argument are in line with that of Ref.\cite{eguchi}, in which all
convergent terms are dropped and only the leading divergent or least
convergent ones are
retained.
A similar argument was given in
Ref.\cite{hippe} for ignoring several $O(1/N_c)$ terms in the higher
order corrections to the pion radius, in which the same argument was made
on heuristic grounds.   This numerical calculation now validates such heuristic
claims. Thus this ordering  plays an additional role in reordering
the set of  terms that occur beyond the $1/N_c$ expansion.

Obviously, by examining the graphs of Figs.~9 to 11, the full cross-section
should be well approximated by that of the $st$-channel alone.
This is given in Fig.~12, and includes the cross terms.
We also  compare our result with the transition rates for the
processes $u\bar u\rightarrow\pi^+\pi^-$ and $ u\bar
u\rightarrow\pi^0\pi^0$ that proceed via the diagrams of Fig.~1.
   Here the process $u\bar u \rightarrow\pi^+\pi^-$
is the dominant one.   We find, for the range of validity of the model, say
$\sqrt s\le 1$GeV, that the hadronization rate of $u$ and $\bar u$ into three
pions, which is also driven by a two pion hadronization graph, is of the
same order of magnitude as $u\bar u$ into two pions, despite the fact that
the phase space is altered.    This result indicates that transport or
other models that include hadronization as a two meson level, require
hadronization into at least three pions also.

%
%

\section{Summary and Conclusions}
In this paper, we have studied the three meson hadronization process
$q\bar q\rightarrow MM'M''$ within the two flavor Nambu--Jona-Lasinio model.
Our motivation stems from the fact that pions are light,
almost massless particles and can thus be produced copiously. 
We have calculated transition rates for this process in order to assess
the importance of this process in comparison with the standard binary
processes that are usually input into transport models.   Since the
NJL model is currently under study as a transport model for a chiral theory
\cite{ogu,aich,greiner}, such questions can be answered within the context
of the model itself.
It has thus been our aim to investigate the transition rate for this process,
in comparison with similar transition rates for
two meson production $\bar q\rightarrow MM'$ for a specific initial state
$q$ and $\bar q$ being $u$ and $\bar u$.

In our analytic discussion,
we have found that there are three types of diagrams
that contribute  in the lowest
level $1/N_c$ expansion that is appropriate to this model.   Our numerical
calculation however 
indicates that an additional criterion for deciding on which
graphs dominate the set of diagrams in a particular
 $1/N_c$ class is the number of
internal fermion lines when a fermion loop diagram is present.   This 
criterion is in
accord with the heuristic methods of Ref.\cite{eguchi}, and which was used
successfully in identifying the logarithmic contribution to the pion radius
\cite{hippe}, and thereby
recovering the known results of sigma models, other chiral models and chiral 
perturbation theory (CHPT)
for this quantity.    We have thus explicitly been able in this paper,
to demonstrate that this crucial heuristic argument given in \cite{hippe}
is in fact valid, in particular in the context of a different calculation.
In
view of the fact that the dominance of
the most divergent graph gives the expected 
 result for the pion radius,  our result here may 
not simply be a  model artifact.    We have at present no way of
knowing how this result would compare with that calculated
within another framework, say a confinement model \cite{koffer}
or within QCD itself, where the mechanism giving rise to the
leading term may be different.     We avoid any speculation here.

   We conclude by stating that the
 numerical calculation shows that our results lie within the same
order of magnitude as the hadronization rates for two pions, thereby
indicating that such processes must necessarily be taken into account in
hadronization at low energies.

\section{Acknowledgments}

We wish to thank our colleagues in the Institute for useful consversations,
in particular, Hilmar Forkel.
This work has been supported  by the Deutsche Forschungsgemeinschaft
DFG under the contract number Hu 233/4-4, and by the German Ministry for
Education and Research (BMBF) under contract number 06 HD 742.

\begin{appendix}

\section{Pauli-Villars regularization for the integrals $I_n$}

The evaluation of $I_n(a_1\dots, a_n)$ as defined in Eq.(\ref{e:in}) is done
using the Pauli-Villars regularization scheme.    The reader is referred to
one of the many papers involving such integrals for $I_1$ and $I_2$, see
for example \cite{dmitra,paul,rehkl}.    We simply quote the final results,
\begin{equation}
I_1= -\frac{i}{(4\pi)^2}[m^2\ln \left(1-(
\frac{\Lambda^2}{m^2+\Lambda^2})^2\right)
+ 2\Lambda^2\ln\left(1+\frac{\Lambda^2}{m^2+\Lambda^2}\right)].
\label{e:i1}
\end{equation}
and $I_2(p,0)=I_2(p)$ is
\begin{equation}
I_2(p) = -\frac i{(4\pi)^2} \sum_{j=0}^2 C_j\int_0^1 dz \ln(m_j^2 - z(1-z)
p^2).
\label{e:i2final}
\end{equation}
where $C_0=1$, $C_1=1$ and $C_2=-2$ is a standard set of coefficients
fulfilling $\sum_j C_j=\sum_jC_jm_j^2=0$.    Here $m_j^2 = m^2+\alpha_j^2
\Lambda^2$, with $\alpha_0=0$, $\alpha_1=2$ and $\alpha_2=1$.
Reference \cite{hooft} gives
\begin{equation}
I_3(p,k) = -\frac{i}{8\pi^2}\sum_{j=0}^2 C_j \int_0^1 dy y\int_0^1 dx
[\frac 1{a^2-b_j} + \frac{a^2-b_j}{2(a^2-b_j)^2}],
\label{e:ipk}
\end{equation}
with $a=y(1-x)p + (1-y)k$ and $b_j=y(1-x)p^2 + (1-y)k^2-m_j^2$, and
\begin{equation}
I_4(p,k,l) = -\frac{3i}{(4\pi)^2}\sum_{j=0}^2C_j\int_0^1 dz z
\int_0^1 dy \int_0^1 dx x^2 [-\frac 1
{4(a^2-b_j)^2} + \frac{a^2-b_j}{6(a^2-b_j)^3}],
\label{e:i3}
\end{equation}
with $a=zx(1-y)p + x(1-z)k + (1-x)l$ and
$b_j = zx(1-y) p^2 + x(1-z) k^2 + (1-x) l^2 - m_j^2$.
An analytical expression for $I_3$ can be found in \cite{hooft}, while
$I_4$ has been evaluated numerically.

\section{Mixed terms in the scattering amplitude}

There are many mixed terms that occur between the various channels.   Here
we list only those that directly enter into the calculation of the
transition rate for
$u\bar u\rightarrow \pi^+\pi^-\pi^0$.

\subsection{Mixed terms between $t$-like graphs}

In this subsection, we list the cross terms that arise
between $t$-like graphs in constructing the
averaged matrix element, and which are required by virtue of
Eq.(\ref{e:mtcontribs}).
    There are three such terms.   The first of these is
%
\begin{eqnarray}
\overline{{\cal M}^{(t,w)}{\cal M}^{(v,y)*}}&=&\frac{f_{t,w}f_{v,y}g_{\pi
qq}^6}
                                                {4N_c}
\left\{-\frac{1}{y-m^2}-\frac{1}{t-m^2}-\frac{(u-m^2)+(x-m^2)}{(v-m^2)(w-m^2)}
                                                   \right.\nonumber\\
&+&           \frac{(t-m^2)(x-m^2)}{(v-m^2)(w-m^2)(y-m^2)}
             +\frac{(u-m^2)(y-m^2)}{(t-m^2)(v-m^2)(w-m^2)}\nonumber\\
&+&     m_{\pi}^2 \left[ \frac{3}{(v-m^2)(w-m^2)} +\frac{1}{(v-m^2)(y-m^2)}
                       +\frac{1}{(t-m^2)(v-m^2)} \right.\nonumber\\
&+&\left.               \frac{-2(t-m^2)-(x-m^2)}{(v-m^2)(w-m^2)(y-m^2)}
                       +\frac{y-x-2(u-m^2)}{(t-m^2)(v-m^2)(w-m^2)}
                                                       \right]\nonumber\\
&+&\left.
m_{\pi}^6\frac{2}{(t-m^2)(v-m^2)(w-m^2)(y-m^2)}\right\}\label{twvy},
\end{eqnarray}
the second that is required is
\begin{eqnarray}
\overline{{\cal M}^{(t,w)}{\cal M}^{(t,y)*}}&=&\frac{f_{t,w}f_{t,y}g_{\pi
qq}^6}
                                                {4N_c}
\left\{-\frac{2}{t-m^2}+\frac{(u-m^2)}{(t-m^2)(w-m^2)}\right.\nonumber\\
&&           +\frac{(v-m^2)}{(t-m^2)(y-m^2)}
             -\frac{(x-m^2)}{(w-m^2)(y-m^2)}\nonumber\\
&&    +m_{\pi}^2 \left[-\frac{1}{(t-m^2)(w-m^2)} -\frac{1}{(t-m^2)(y-m^2)}
                       +\frac{2}{(t-m^2)^2} \right.\nonumber\\
&&\left.               +\frac{2}{(w-m^2)(y-m^2)}
                       +\frac{(x-m^2)}{(t-m^2)(w-m^2)(y-m^2)}
                                                       \right]\nonumber\\
&&\left. -m_{\pi}^6\frac{2}{(t-m^2)^2(w-m^2)(y-m^2)}  \right\}\label{twty}
\end{eqnarray}
and the last one is
\begin{eqnarray}
\overline{{\cal M}^{(v,y)}{\cal M}^{(t,y)*}}&=&\frac{f_{v,y}f_{t,y}g_{\pi
qq}^6}
                                                {4N_c}
\left\{-\frac{2}{y-m^2}+\frac{(w-m^2)}{(t-m^2)(y-m^2)}\right.\nonumber\\
&+&                       \frac{(x-m^2)}{(v-m^2)(y-m^2)}
                         -\frac{(u-m^2)}{(t-m^2)(v-m^2)}\nonumber\\
&+&    m_{\pi}^2 \left[ \frac{2}{(y-m^2)^2} +\frac{1}{(t-m^2)(v-m^2)}
                  +\frac{2u-w-x}{(t-m^2)(v-m^2)(y-m^2)}\right]\nonumber\\
&&\left. -m_{\pi}^6\frac{2}{(t-m^2)(v-m^2)(y-m^2)^2}  \right\} .\label{vyty}
\end{eqnarray}

\subsection{Mixed terms between $s$-like graphs}

Mixed terms between $s$-like graphs do not take a simple form in terms
of the extended Mandelstam-like variables.   Generally, one has
\begin{eqnarray}
&&{\cal M}_\pi(p_a,p_b,p_c){\cal M}^*_\pi(p_a',p_b',p_c') \nonumber \\
&&\hspace{2cm} = 8G^2 g_{\pi q q}^2 (N_f^j)^2 \delta_{c_1c_2} \frac
{|\bar v_2 i\gamma_5 u_1|^2 A_{\pi\pi\pi\pi}(p_a,p_b,p_c) A^*_{\pi\pi\pi\pi}
(p_a',p_b',p_c')}
{|1-2G\Pi_{PS}(k^2)|^2},
\label{e:mix1}
\end{eqnarray}
where $(p_a,p_b,p_c)$ or $(p_a',p_b',p_c')$  represent the combinations
$(p_3,p_4,p_5)$, $(p_4,p_5,p_3)$ or $(p_5,p_3,p_4)$.   Averaging over
incoming states and summing over final states leads to
\begin{eqnarray}
&&\overline { {\cal M}_\pi(p_a,p_b,p_c){\cal M}^*_\pi(p_a',p_b',p_c') }
 \nonumber \\
&&\hspace{2cm}= \frac {2G^2g_{\pi q q}^6 (N_f^j)^2}{N_c}
\frac
{4s A_{\pi\pi\pi\pi}(p_a,p_b,p_c) A^*_{\pi\pi\pi\pi}
(p_a',p_b',p_c')}
{|1-2G\Pi_{PS}(k^2)|^2},
\label{e:mix2}
\end{eqnarray}
which is evaluated directly numerically.

\subsection{Mixed terms between $s$-like and $t$-like graphs}

Here three such products are required.  The first of these is
the averaged matrix element of ${\cal M}_\pi(p_a,p_b,p_c)$ together
with its complex conjugate, which can
be constructed as
\begin{eqnarray}
&&\overline{{\cal M}_\pi(p_a,p_b,p_c) {\cal M}^{(t,w)*} }  \nonumber \\
&&\hspace{2cm} = \frac 1{N_c}\sum_{s_1,s_2} \bar v_2 i\gamma_5 u_1
\frac{4iG g_{\pi q q}^3 N_f^j
A_{\pi\pi\pi\pi}(p_a,p_b,p_c)}{1-2G\Pi_{PS}(k^2)} \frac {-g_{\pi q q}^3
f_{t,w}}{(t-m^2)(w-m^2)} \nonumber \\
&&\hspace{2.5cm}
\times \bar u_1\gamma_5T_5(\not \! p_1 - \not \! p_3 + m)\gamma_5 T_4
(\not\! p_1 - \not \! p_3 - \not \! p_4 + m) \gamma_5 T_3 v_2.
\label{e:mix3}
\end{eqnarray}
Upon taking the traces and forming the complex conjugate, one finds the
expression
\begin{eqnarray}
&&\overline{{\cal M}_\pi(p_a,p_b,p_c) {\cal M}^{(t,w)*} }
+ \overline{{\cal M}^{(t,w)}{\cal M}^*_\pi(p_a,p_b,p_c)} = \nonumber \\
&& -\frac{Gg_{\pi q q}^6 f_{t,w} N_f^j}{N_c(t-m^2)(w-m^2)}
2{\it Re} \left(\frac{A_{\pi\pi\pi\pi}(p_a,p_b,p_c)}{1-2G\Pi_{PS}(k^2)}
\right) \{(v-m^2-m_\pi^2)(w-m^2-m_\pi^2) \nonumber \\
&&(t-m^2-m_\pi^2)[(2(w-m^2) + (x-m^2-2m_\pi^2) - (u-m^2-m_\pi^2
)(y-m^2-2m_\pi^2)\}, \nonumber \\
\end{eqnarray}
and correspondingly
\begin{eqnarray}
&&\overline{{\cal M}_\pi(p_a,p_b,p_c) {\cal M}^{(v,y)*} }
+ \overline{{\cal M}^{(v,y)}{\cal M}^*_\pi(p_a,p_b,p_c)} = \nonumber \\
&& -\frac{Gg_{\pi q q}^6 f_{v,y} N_f^j}{N_c(t-m^2)(w-m^2)}
2{\it Re} \left(\frac{A_{\pi\pi\pi\pi}(p_a,p_b,p_c)}{1-2G\Pi_{PS}(k^2)}
\right) \{(u-m^2-m_\pi^2)(y-m^2-m_\pi^2) \nonumber \\
&&+(v-m^2-m_\pi^2)[(2(y-m^2) + (w-m^2-2m_\pi^2)]-(t-m^2-m_\pi^2)(x-m^2
-2m_\pi^2)\}, \nonumber \\
\end{eqnarray}
and
\begin{eqnarray}
&&\overline{{\cal M}_\pi(p_a,p_b,p_c) {\cal M}^{(t,y)*} }
+ \overline{{\cal M}^{(t,y)}{\cal M}^*_\pi(p_a,p_b,p_c)} = \nonumber \\
&& -\frac{Gg_{\pi q q}^6 f_{v,y} N_f^j}{N_c(t-m^2)(y-m^2)}
2{\it Re} \left(\frac{A_{\pi\pi\pi\pi}(p_a,p_b,p_c)}{1-2G\Pi_{PS}(k^2)}
\right) \{ (u-m^2-m_\pi^2)(y-m^2-m_\pi^2) \nonumber \\
&&+ (t-m^2-m_\pi^2)[(2(y-m^2) + (x-m^2-2m_\pi^2)] -
(v-m^2-m_\pi^2)(w-m^2-2m_\pi^2) \}
\nonumber \\
\end{eqnarray}
which ends this subsection.

\subsection{Mixed terms between $st$-like and $t$-like graphs}

Using the generic labels $p_A$, $p_B$ and $p_C$ to denote the outgoing
pions in the $t$-like channel, one is required to form the cross term
between the expression
\begin{equation}
{\cal M}^{t-{{\rm like}}} = (ig_\pi)^2 f\bar v(p_2) i\gamma_5
\frac{ i(\not p_C -
p_2+m)}{(p_C-p_2 )^2-m^2} i\gamma_5\frac{i(p_1-p_A+m)}{(p_1-p_A)^2-m^2}
i\gamma_5 u(p_1),
\label{e:ABC}
\end{equation}
where $f$ is a flavor factor, and the expression for ${\cal M}^{st-{{\rm
like}}}$ in Eq.(\ref{e:missing}).   The averaged cross product using the
first term of Eq.(\ref{e:missing}) gives
\begin{eqnarray}
&&\overline{{\cal M}_1^{(st)-{{\rm like}}} {\cal M}^{t-{{\rm like}*}} }
= \frac{1}{4N_c}\frac{i g_{\pi q q}^4 A_{\sigma \pi\pi} f}
{(w-m^2)[(p_1-p_A)^2-m^2][(p_C-p_2)^2-m^2]} \nonumber \\
&& \hspace{0.1in} {\rm tr}_\gamma [(\not\! p_1 + m)\gamma^5(\not\! p_1 - \not
\! p_A +m)
\gamma^5(\not \! p_C - \not\!p_2 +m)\gamma^5(\not\!p_2-m) \gamma^5
(\not\!p_5-\not \! p_2 + m)].
\label{e:long}
\end{eqnarray}
The spinor trace can be performed, so that
\begin{eqnarray}
&&\overline{{\cal M}_1^{(st)-{{\rm like}}} {\cal M}^{t-{{\rm like}*}} }
= \frac{ig_{\pi q q}^4 A_{\sigma\pi\pi} fm}{N_c (w -m^2)}
\frac1{[(p_1-p_A)^2-m^2][(p_C-p_2)^2 -m^2]} \nonumber \\
&& \hspace{0.2in}
\times[-(p_A\cdot p_C)p_5\cdot(p_1+p_2) + (p_A\cdot p_5) (p_1-p_2)\cdot
p_C - (p_C\cdot p_5) p_A\cdot(p_1-p_2)]. \nonumber \\
\label{e:mstts}
\end{eqnarray}
Carrying out this procedure with the second term of Eq.(\ref{e:missing})
leads to precisely the same spinor trace, so that the expression differs
only from the above in that $\omega - m^2$ is replaced by $v-m^2$, in
building
$\overline{{\cal M}_2^{(st)-{{\rm like}}} {\cal M}^{t-{{\rm like}*}} }$
and
$\overline{{\cal M}^{(st)-{{\rm like}}} {\cal M}^{t-{{\rm like}*}} }
= \overline{{\cal M}_1^{(st)-{{\rm like}}} {\cal M}^{t-{{\rm like}*}}
+
{\cal M}_2^{(st)-{{\rm like}}} {\cal M}^{t-{{\rm like}*}} }$.

When the incoming states are $u$ and $\bar u$, and the outgoing states
$\pi^+\pi^-\pi^0$, three $t$-like graphs can be constructed in which the
outgoing momenta are permuted.   These cases are
(a) $p_A=p_5=p_{\pi^0}$,  $p_B=p_3=p_{\pi^+}$ and  $p_C=p_4=p_{\pi^-}$
with $f=2$; (b)  $p_A=p_3=p_{\pi^+}$,  $p_B=p_5=p_{\pi^0}$
and  $p_C=p_4=p_{\pi^-}$ with $f=-2$, and (c)  $p_A=p_3=p_{\pi^+}$,
 $p_A=p_4=p_{\pi^-}$ and  $p_C=p_5=p_{\pi^0}$ with $f=2$.
Combining these three contributions and moving to the extended Mandelstam
variables, one finds the form
\begin{eqnarray}
\overline{{\cal M}^{(st)-{{\rm like}}} {\cal M}^{t-{{\rm like}*}} }
&=&  \frac{img_{\pi q q}^4 A_{\sigma \pi\pi}}{N_c} \left[\frac 1{w-m^2}
+ \frac 1{v-m^2}\right] \nonumber \\
&\times& \large  [ \frac{v-m^2)((x-u-v-m^2)
-m_\pi^2(x-y-v-m^2)}{[v-m^2][y-m^2]} \nonumber \\
&+& \frac {(w-m^2)(v-y) - (v-m^2)(t-m^2) -m_\pi^2(w-t-u+m^2)}
{[t-m^2][y-m^2]} \nonumber \\
&+& \frac{(w-m^2)(y-t-v+m^2) - m_\pi^2(u-v-t+m^2)}{[t-m^2][w-m^2]}\large ]
\label{e:monster2}
\end{eqnarray}
in total.  The mixed term between the $(st)$-like and $s$-like graphs
does not have a simple form, and must be constructed from
Eq.(\ref{e:missing}) and Eq.(\ref{e:maverage}).

\end{appendix}

\newpage
\begin{center}
FIGURE CAPTIONS.
\end{center}

\noindent
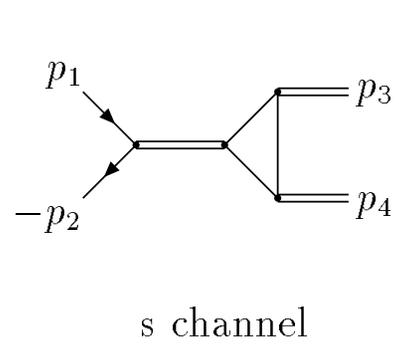
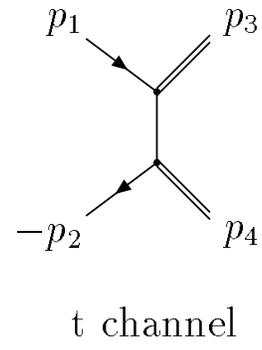
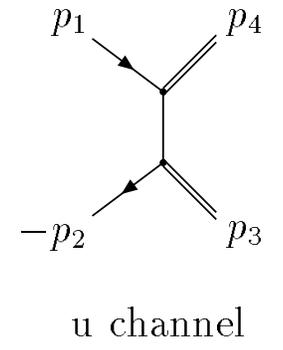
FIG.1:  $s$, $t$ and $u$ channel graphs available for the
hadronization of $q\bar q\rightarrow MM'$, to lowest order in
the $1/N_c$ expansion.

\vskip 0.1in
\noindent
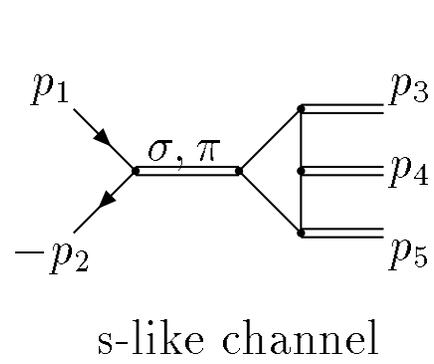
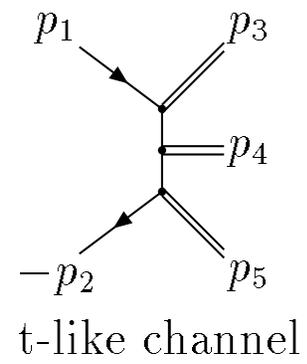
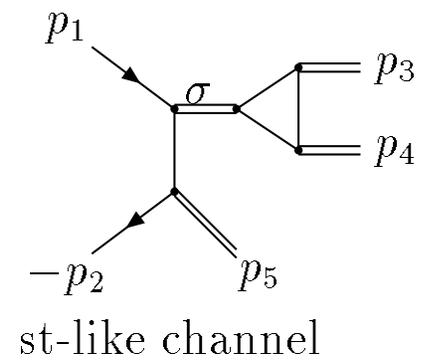
FIG.2:   $s$-like, $t$-like and $st$-like  channels available for the
hadronization of $q\bar q\rightarrow MM'M''$, to lowest order in
the $1/N_c$ expansion.

\vskip 0.1in
\noindent
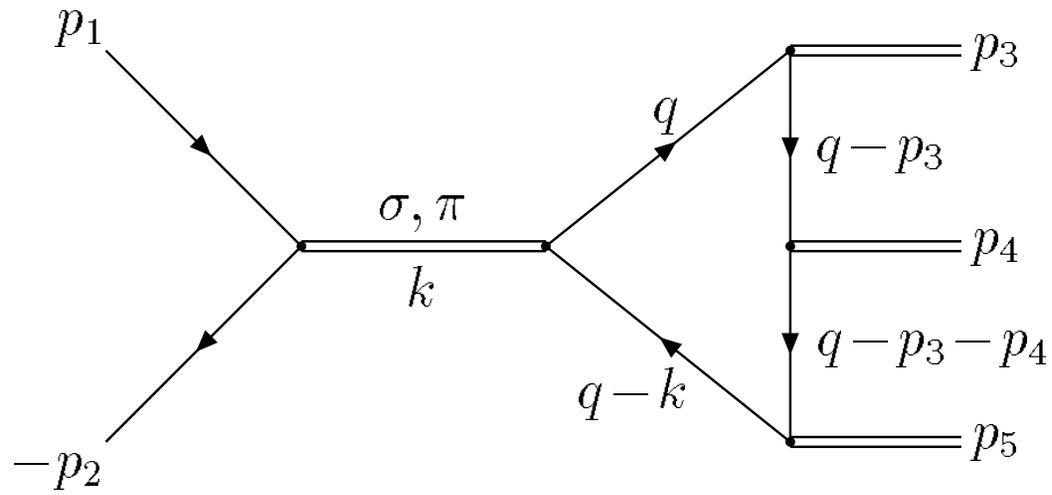
FIG.3:  Exchange graph in the $s$-like channel, for
$q\bar q\rightarrow MM'M''$.

\vskip 0.1in
\noindent
FIG.4:   Vertex associated with the $s$-like channel hadronization
graph.

\vskip 0.1in
\noindent
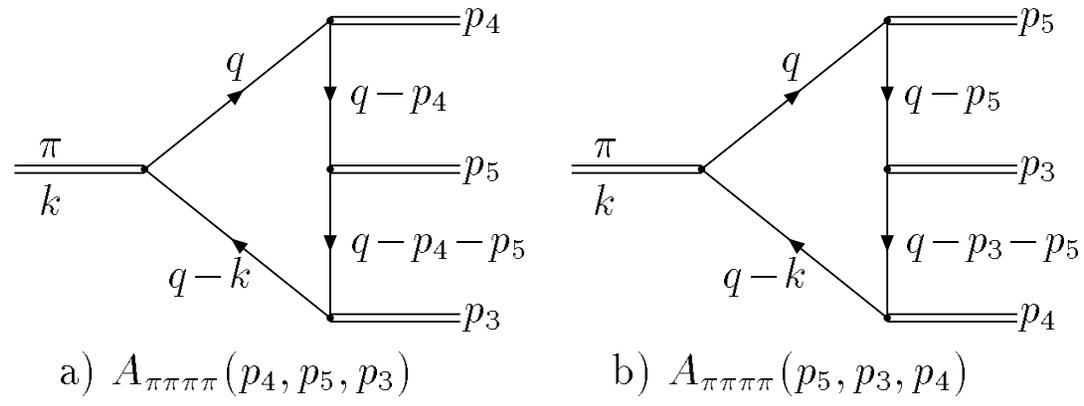
FIG.5:  The vertex $\pi\rightarrow\pi\pi\pi$ cross
 channels obtained by permutations of the external
meson indices.

\vskip 0.1in
\noindent
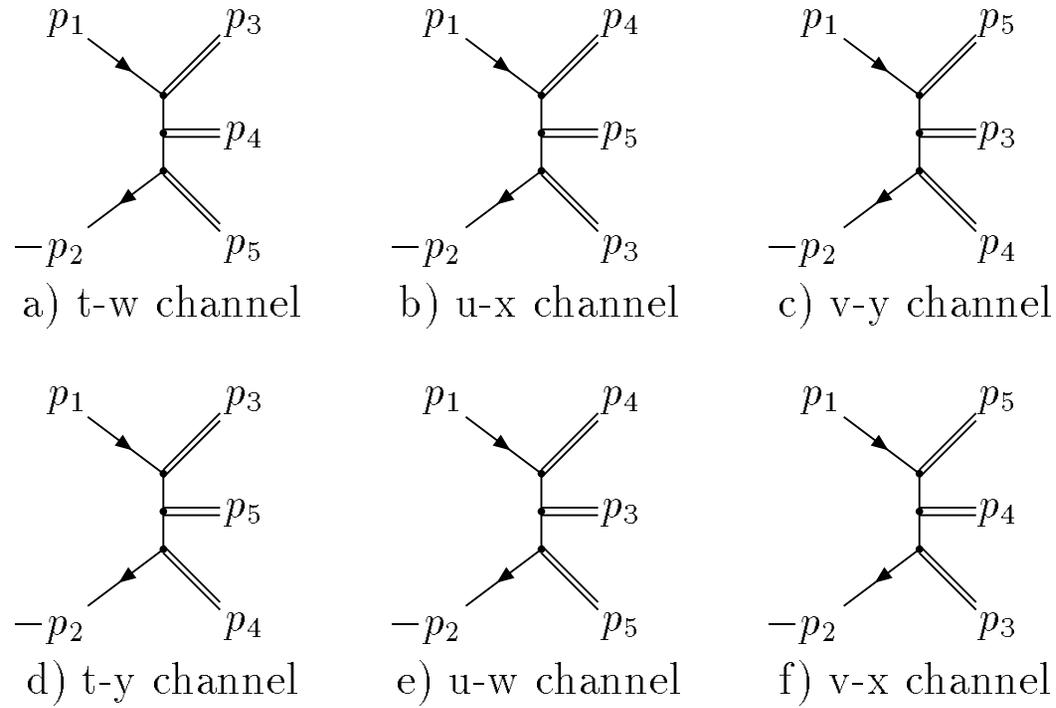
FIG.6:  Six available $t$-like channels.

\vskip 0.1in
\noindent
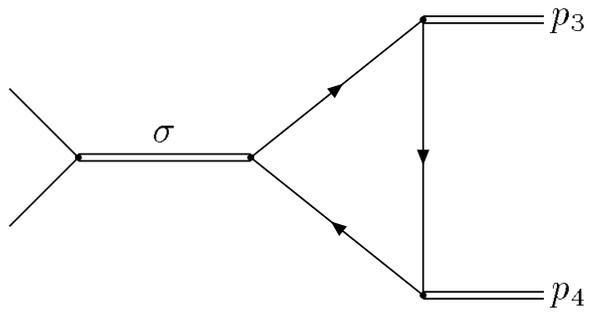
FIG.7:  Two particle production from an intermediate state.

\vskip 0.1in
\noindent
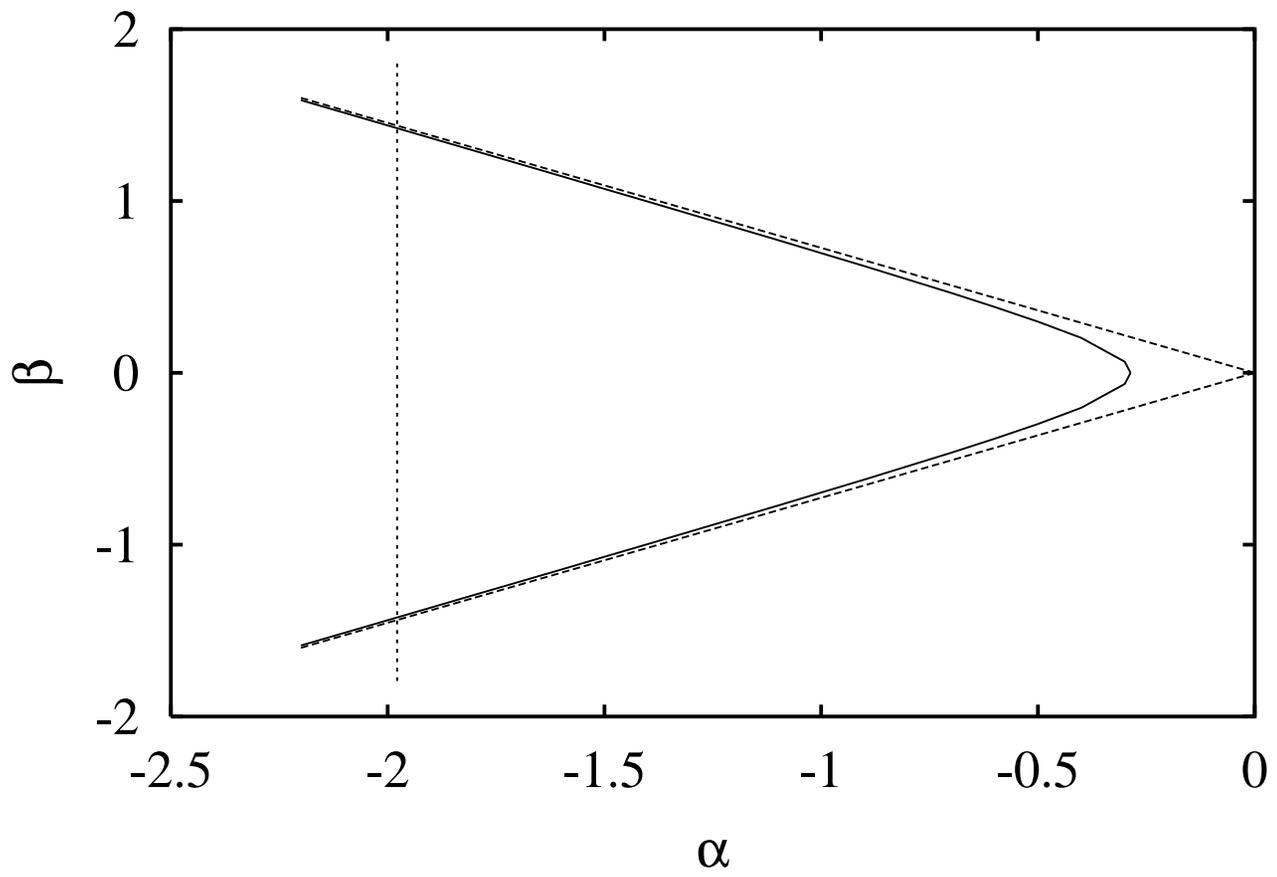
FIG.8:  Region of integration in the $\alpha$-$\beta$ plane.
The solid line indicates the negative hyperbola that is determined by
Eq.(\ref{e:hyper}).   The dotted vertical line is the boundary given by
Eq.(\ref{e:vertbound}).   The asymptotes of the hyperbola are indicated by the
dashed
line.  The surface $F_1$ is bounded to the left of the solid line and the
right of the dotted vertical line.   $W_1$ is the triangle bounded to the
right by the dotted vertical line and to the left by the
dashed lines.

\vskip 0.1in
\noindent
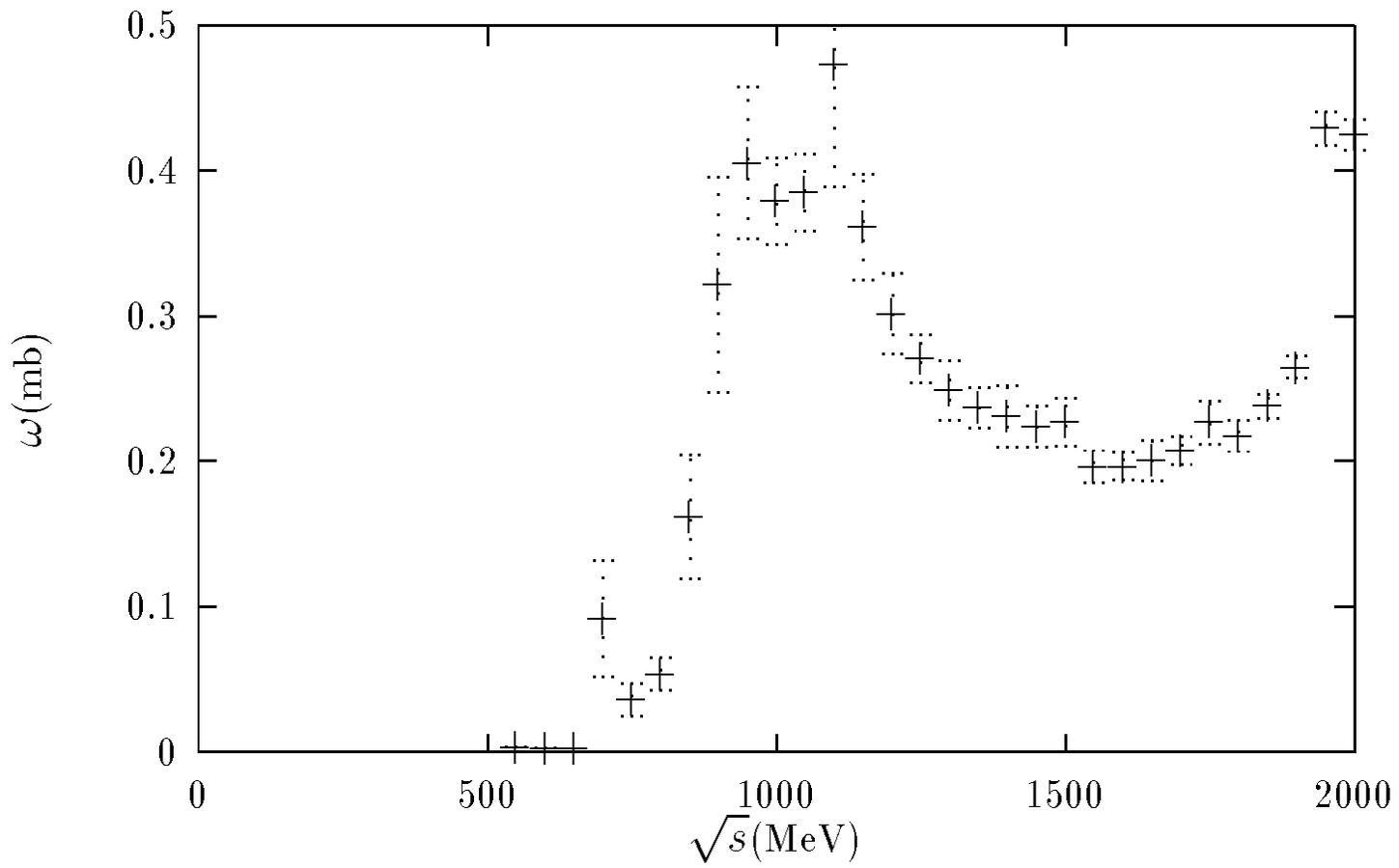
FIG.9:   Transition rate $\omega$ plotted as a function of $\sqrt
s$ for the $s$-like channel.

\vskip 0.1in
\noindent
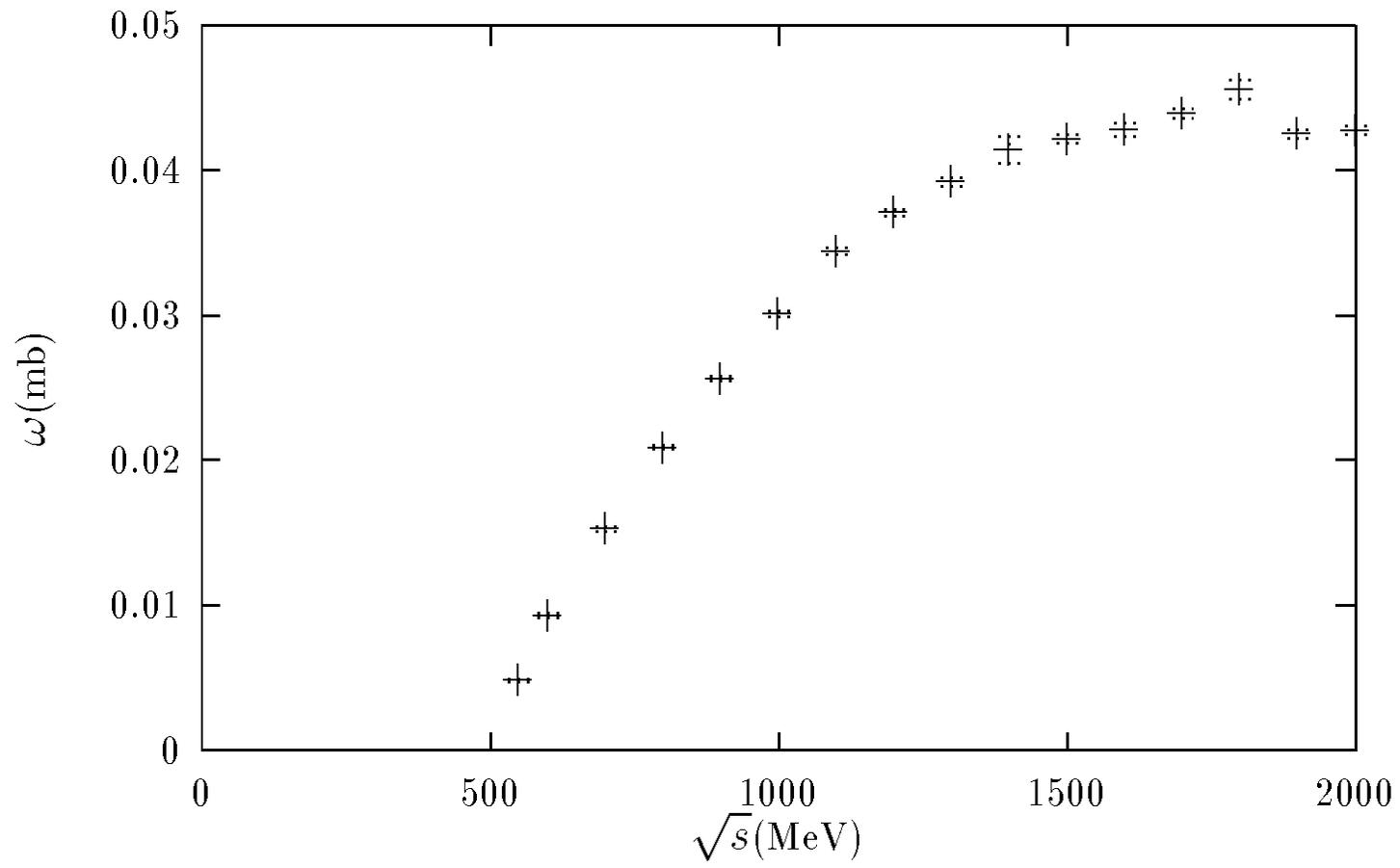
FIG.10:  Transition rate $\omega$ plotted as a function of $\sqrt
s$ for the $t$-like channel.

\vskip 0.1in
\noindent
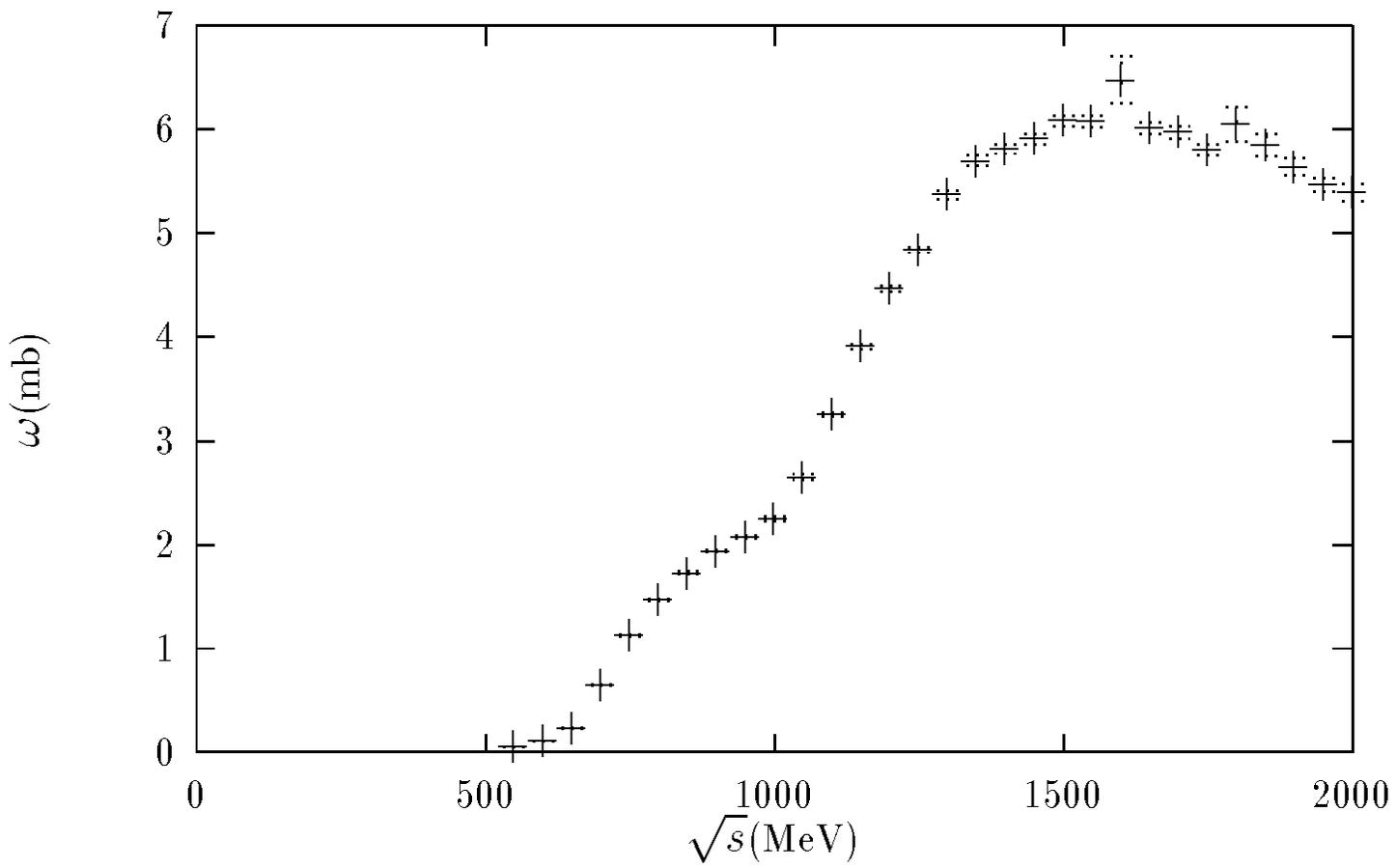
FIG.11:   Transition rate $\omega$ plotted as a function of $\sqrt
s$ for the $st$-like channel.

\vskip 0.1in
\noindent
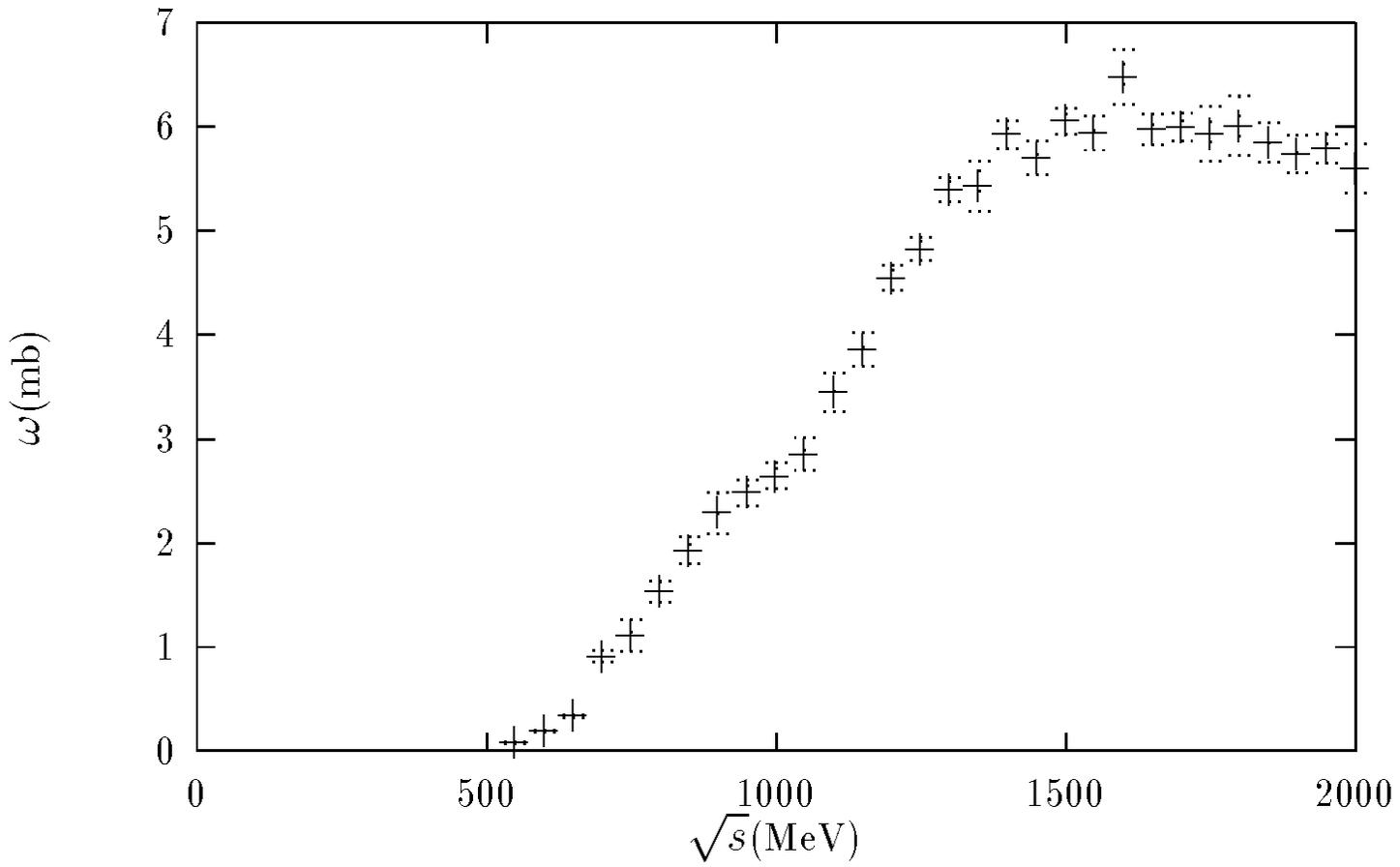
FIG.12:  Transition rate $\omega$ plotted as a function of $\sqrt
s$.

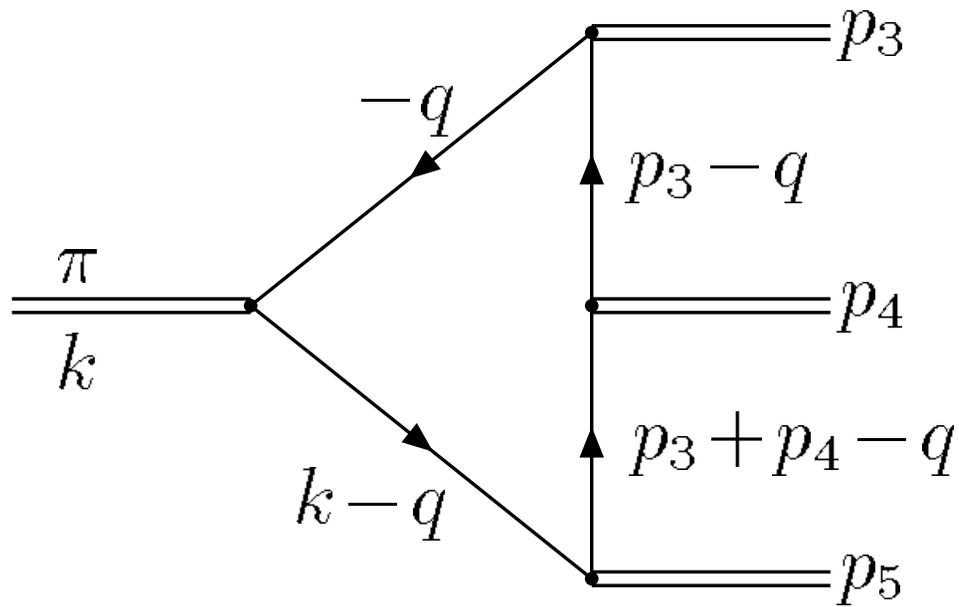
\end{document}